\documentclass[sigconf,screen=true,natbib=false]{acmart}
\usepackage{amsmath,amsfonts}
\usepackage{multirow}
\usepackage{multicol}
\usepackage{amsthm}
\usepackage{listings}
\usepackage{algorithm}
\usepackage{algpseudocode}
\usepackage{graphicx}
\usepackage{textcomp}
\usepackage{xcolor}
\usepackage{cleveref}
\usepackage{booktabs}
\usepackage{comment}
\usepackage{tabularx}
\usepackage{multirow}
\usepackage{bm}
\usepackage{url}
\usepackage{xspace}
\usepackage{pifont}
\usepackage{arydshln}
\usepackage{verbatim}
\usepackage[referable]{threeparttablex}
\usepackage{calc} 
\usepackage{color}
\usepackage{float}
\usepackage{subcaption}
\usepackage[utf8]{inputenc}
\usepackage{booktabs} 
\usepackage{tabularx} 
\usepackage{tikz}
\usetikzlibrary{shapes.geometric, arrows}
\tikzstyle{startstop} = [rectangle, rounded corners, minimum width=3cm, minimum height=1cm,text centered, draw=black]
\tikzstyle{io} = [trapezium, trapezium left angle=70, trapezium right angle=110, minimum width=3cm, minimum height=1cm, text centered, draw=black]
\tikzstyle{process} = [rectangle, minimum width=3cm, minimum height=1cm, text centered, draw=black]
\tikzstyle{arrow} = [thick,->,>=stealth]

\usepackage{filecontents}                                  
\usepackage{pgfplots}
\usepackage{pgfplotstable}
\usepackage{scalefnt}
\pgfplotsset{compat=newest}

\definecolor{USTgold}{RGB}{153,102,0}
\definecolor{USTyellow}{RGB}{204,153,0}
\definecolor{USTyellowlight}{RGB}{255,212,0}
\definecolor{USTorange}{RGB}{255,166,26}
\definecolor{USTpink}{RGB}{255,157,157}
\definecolor{USTblue}{RGB}{0,51,102}
\definecolor{USTmiddle}{RGB}{0,116,188}
\definecolor{USTlight}{RGB}{99,202,225}
\definecolor{USTgray}{RGB}{204,204,204}
\definecolor{USTred}{RGB}{237,27,47}
\definecolor{USTdarkred}{RGB}{124,35,72}

\definecolor{CUHKorange}{RGB}{244,106,18} 
\definecolor{CUHKblue}{RGB}{0,111,190}    
\definecolor{CUHKgreen}{RGB}{0,127,128}   
\definecolor{CUHKred}{RGB}{228,46,36}     
\definecolor{CUHKyellow}{RGB}{198,148,34} 
\definecolor{CUHKdark}{RGB}{114,44,114}   
\definecolor{CUHKmiddle}{RGB}{144,44,144} 
\definecolor{CUHKlight}{RGB}{167,44,167}

\iftrue
\def\BibTeX{{\rm B\kern-.05em{\sc i\kern-.025em b}\kern-.08em
    T\kern-.1667em\lower.7ex\hbox{E}\kern-.125emX}}

\fi

\acmConference[DAC '24]{Design Automation Conference}{June 23--27, 2024}{San Francisco, CA}

\iftrue
\fi

\iftrue
\usepackage{titlesec}
\titlespacing\section{2pt}{5pt plus 1pt minus 1pt}{0pt plus 1pt minus 1pt}
\titlespacing\subsection{2pt}{5pt plus 1pt minus 1pt}{0pt plus 1pt minus 1pt}
\titlespacing\subsubsection{2pt}{5pt plus 1pt minus 1pt}{2pt plus 1pt minus 1pt}
\usepackage[inline]{enumitem}
\setlist{leftmargin=4.08mm}
\fi

\iftrue
\fi

\newcommand\blfootnote[1]{%
  \begingroup
  \renewcommand\thefootnote{}\footnote{#1}%
  \addtocounter{footnote}{-1}%
  \endgroup
}

\usepgfplotslibrary{groupplots}

\copyrightyear{2024}
\acmYear{2024}
\setcopyright{acmlicensed}\acmConference[ICCAD '24]{IEEE/ACM International Conference on Computer-Aided Design}{October 27--31, 2024}{New York, NY, USA}
\acmBooktitle{IEEE/ACM International Conference on Computer-Aided Design (ICCAD '24), October 27--31, 2024, New York, NY, USA}
\acmDOI{10.1145/3676536.3676793}
\acmISBN{979-8-4007-1077-3/24/10}

\begin{document}

\title{UFO-MAC: A Unified Framework for Optimization of High-Performance Multipliers and Multiply-Accumulators}

\author{Dongsheng Zuo, Jiadong Zhu, Chenglin Li, Yuzhe Ma$^{*}$}
\affiliation{
    \institution{
    Microelectronics Thrust \\The Hong Kong University of Science and Technology (Guangzhou) }
}
\email{yuzhema@hkust-gz.edu.cn}

\begin{abstract}
Multipliers and multiply-accumulators (MACs) are critical arithmetic circuit components in the modern era. 
As essential components of AI accelerators, they significantly influence the area and performance of compute-intensive circuits. 
This paper presents UFO-MAC, a unified framework for the optimization of multipliers and MACs. 
Specifically, UFO-MAC employs an optimal compressor tree structure and utilizes integer linear programming (ILP) to refine the stage assignment and interconnection of the compressors. 
Additionally, it explicitly exploits the non-uniform arrival time profile of the carry propagate adder (CPA) within multipliers to achieve targeted optimization.
Moreover, the framework also supports the optimization of fused MAC architectures. 
Experimental results demonstrate that multipliers and MACs optimized by UFO-MAC Pareto-dominate state-of-the-art baselines and commercial IP libraries. 
The performance gain of UFO-MAC is further validated through the implementation of multipliers and MACs within functional modules, underlining its efficacy in real scenarios.

\end{abstract}

\maketitle
\pagestyle{plain}

\section{Introduction}
\blfootnote{*Corresponding author}
%
%
%
%
%


In digital circuit design, multipliers and multiply-accumulators are fundamental arithmetic components,
which are particularly critical for computation-intensive applications. 
Consequently, the optimization of high-performance multipliers and MACs becomes imperative, as their optimization significantly influences overall performance, energy efficiency, and area footprint.


The fundamental architecture of a multiplier typically includes three key components: a partial product generator (PPG), a compressor tree (CT), and a carry propagate adder (CPA). 
The CT efficiently compresses the partial products generated by the PPG into two rows, which are then summed by the CPA to produce the final product.
The CT is crucial for efficiently performing the addition of partial products generated by the PPG in parallel \cite{Datapath-1964TC-Wallace,Datapath-1983ARITH-Dadda}. 
Moreover, there have been numerous customized designs specifically optimized for specific technology nodes and applications \cite{Datapath-1993ASAP-Bickerstaff, Datapath-1993TVLSI-Fadavi-Ardekani,Datapath-2005ISCAS-Itoh,Datapath-2014ATC-Luu}.
While customized designs of multipliers offer precise control, they often lack the flexibility to quickly adapt to new technology nodes and applications. 
To address this, algorithmic methods have emerged as more flexible solutions that leverage advances in algorithmic strategies, mathematical programming, and heuristic search techniques.
The three-dimensional method (TDM) has been introduced for the design of compressor trees\cite{Datapath-1996TC-Oklobdzija,Datapath-1995ARITH-Martel,Datapath-1998TC-Stelling}. 
In FPGA design, integer linear programming (ILP) has been applied effectively to compressor tree optimization, utilizing specialized counter resources to efficiently balance area and delay \cite{Datapath-2008DATE-Parandeh-Afshar}. 
Subsequent enhancements have included sophisticated modeling techniques \cite{Datapath-2014FPL-Kumm,Datapath-2017ARITH-Kumm}, heuristics to refine the solution space \cite{Datapath-2018TC-Kumm}, and the comprehensive global optimization of PPG and CPA \cite{Datapath-2023TC-Böttcher}. 
The ILP for ASIC multiplier optimization was proposed in GOMIL \cite{Datapath-2021DATE-Xiao}, where the ILP was used to minimize the area of the compressor tree, and linear programming was utilized for the optimization of the CPA. 

\begin{figure}[!tb]
\centering
\includegraphics[width=0.98\linewidth]{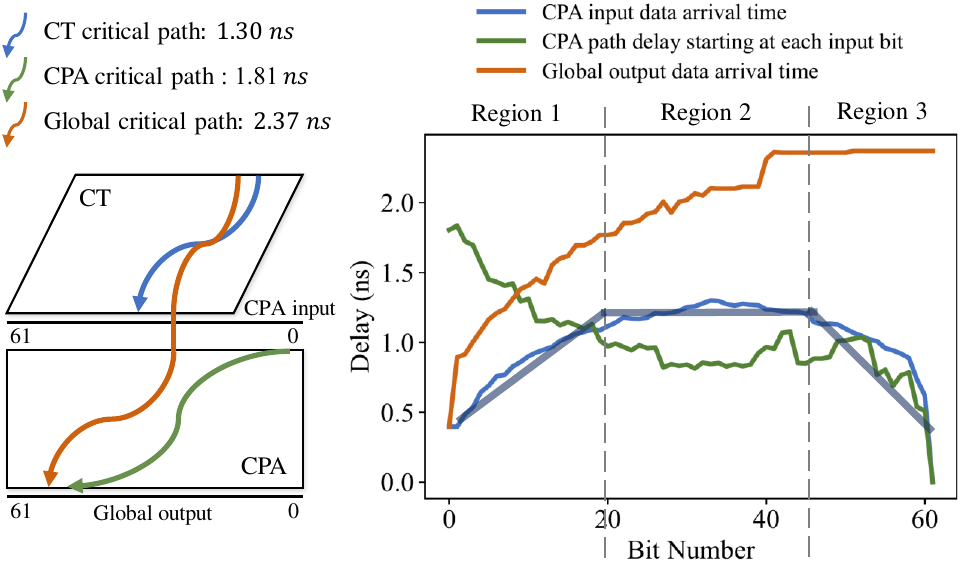}
\vspace{-0.3cm}
\caption{Motivating example: The optimization of the CT and the CPA are not decoupled; CPA exhibits a non-uniform arrival time profile, requiring optimization strategies different from those of traditional adders methodology}
\label{fig:motivation}
\end{figure}

Regarding CPA design, prefix adders are adopted for more efficient addition. 
Prefix adders incorporate regular structures that are optimized regarding logic level, fan-out, and wire tracks, as seen in Sklansky tree \cite{Datapath-1960TC-Sklansky}, Kogge-Stone tree \cite{Datapath-1973TC-Kogge}, and Brent-Kung tree \cite{Datapath-1982TC-Brent}.
Automated synthesis approaches have introduced greater flexibility.
Modify-based methods modify regular structures through equivalent transformations to meet design constraints \cite{Datapath-1990DAC-Fishburn, 1996-IWLAS-Zimmermann, Datapath-2002ISPAN-Fishburn}. 
In addition, ILP has been utilized to systematically explore and optimize adder trees, employing analytical models that account for area, power, and timing \cite{Datapath-2007ASPDAC-Liu}. 
Furthermore, Roy~\textit{ et al.} have advanced this field by proposing an exhaustive search approach that incorporates pruning strategies, which effectively streamline the design process by focusing only on the most promising configurations \cite{Datapath-2013DAC-Roy, Datapath-2014TCAD-Roy, Datapath-2016TACD-Roy}.
Recently, machine learning methodologies have emerged, which employ surrogate evaluators to assess design variants during optimization \cite{DSE-2022TCAD-Geng,DSE-2019TCAD-MA} or train an agent to directly optimize a design \cite{RL-2021DAC-Roy,Datapath-2023DAC-Zuo}.
Notably, reinforcement learning has been applied to refine traditional datapath architectures, such as in PrefixRL~\cite{RL-2021DAC-Roy}, where it optimizes prefix adders by modifying classical adder structures. 
Similarly, RL-MUL \cite{Datapath-2023DAC-Zuo} represents compressor trees as tensors and employs a reinforcement learning agent to optimize multiplier design.
In addition, the interconnect order within the CT may also impact the delay of CT, while RL-MUL only considered searching for the total compressor number in each column of CT.

Despite that each component has been extensively explored in previous studies, obtaining a high-performance multiplier and MAC is still non-trivial today.
On the one hand, the design space of CT in multipliers has not been well explored in prior research. 
The methods for compressor assignment and the interconnection orders between compressors significantly influence CT performance.
These aspects are often overlooked in existing works \cite{Datapath-2021DATE-Xiao,Datapath-2023DAC-Zuo}. 
On the other hand, the three components of multipliers and MACs - PPG, CT, and CPA - are not decoupled.
As illustrated in \Cref{fig:motivation}, the global critical path of a multiplier does not simply accumulate the critical paths of CT and CPA.
On the right side of \Cref{fig:motivation}, we can see that the CT output profile exhibits a ``trapezoidal'' shape, where the data at the least significant bit (LSB) and most significant bit (MSB) arrive first, and the data for the middle bits arrive last.
This can be segmented into three regions, and the observation provides us with two insights:
First, in region 2, where the CT data arrive last, there is a necessity to employ high-speed prefix structures to accommodate the critical path delay. 
Conversely, in regions 1 and 3, where the data arrive earlier, there is no need for fast prefix structures. 
By leveraging the non-uniform arrival profile of regions 1 and 3, we can effectively optimize the area without compromising the performance of the overall design.
Previous work GOMIL \cite{Datapath-2021DATE-Xiao} has focused only on minimizing the area of the compressor tree and the depth of the Carry Propagation Adder (CPA), while not exploiting the non-uniform arrival profile. 
Other studies such as RL-MUL \cite{Datapath-2023DAC-Zuo} concentrated solely on the compressor tree while overlooking the significant impact that CPA optimizations can have on the overall performance of multipliers. 

In contrast, a more effective strategy involves targeted optimizations of the CPA based on the CT output profile. 
While there are existing works on non-uniform arrival adders, such as the hybrid adder using a carry skip adder \cite{Datapath-1995Asilomar-Stelling, Datapath-1995TVLSI-Oklobdzija, Datapath-2012TODAES-Kim}, and approaches that transform non-uniform arrival times into logic depth constraints for prefix graphs \cite{1996-IWLAS-Zimmermann,Datapath-2007GLSVLSI-Matsunaga}.
However, logic depth provides a low fidelity of path delay, and node fanout can significantly impact path delay\cite{Datapath-2014TCAD-Roy}, which is not considered in these approaches.
To address these limitations, we propose UFO-MAC, a unified framework for the optimization of high-performance multipliers and MACs.
UFO-MAC not only adopts area-optimal CTs but also expands the design space to utilize ILP to optimize compressor assignment and interconnection orders, which ensures effective area and delay optimization.
For CPA design, UFO-MAC explicitly leverages the non-uniform input arrival profile, adopting a linear timing model that accounts for both fanout and logic depth. 
This model provides a higher fidelity that guides the CPA optimization more effectively. 
Starting from an area-efficient initial CPA structure, the framework applies depth and fanout optimization to meet timing constraints, thereby enhancing the overall performance of the adder.

In summary, the contributions of UFO-MAC are as follows:
\begin{itemize}
    \item We propose UFO-MAC, a unified framework for the optimization of multipliers and MACs, enhancing both area and delay metrics.
    \item We introduce area-optimal CT structures and extend the design space to optimize the interconnect order of the compressor trees.
    \item We explicitly explore the non-uniform arrival profile for targeted optimization of CPAs based on our max-path fanout timing model.
    \item Experimental results confirm that UFO-MAC optimized multipliers and MACs exceed all baseline designs. The effectiveness of these optimized designs has been further validated in practical applications, including signal processing and AI acceleration.
\end{itemize}

\section{Preliminaries}

\subsection{Multiplier Architecture}
\label{subsec:mult-arch}
\begin{figure}[!tb]
\centering
\includegraphics[width=.768\linewidth]{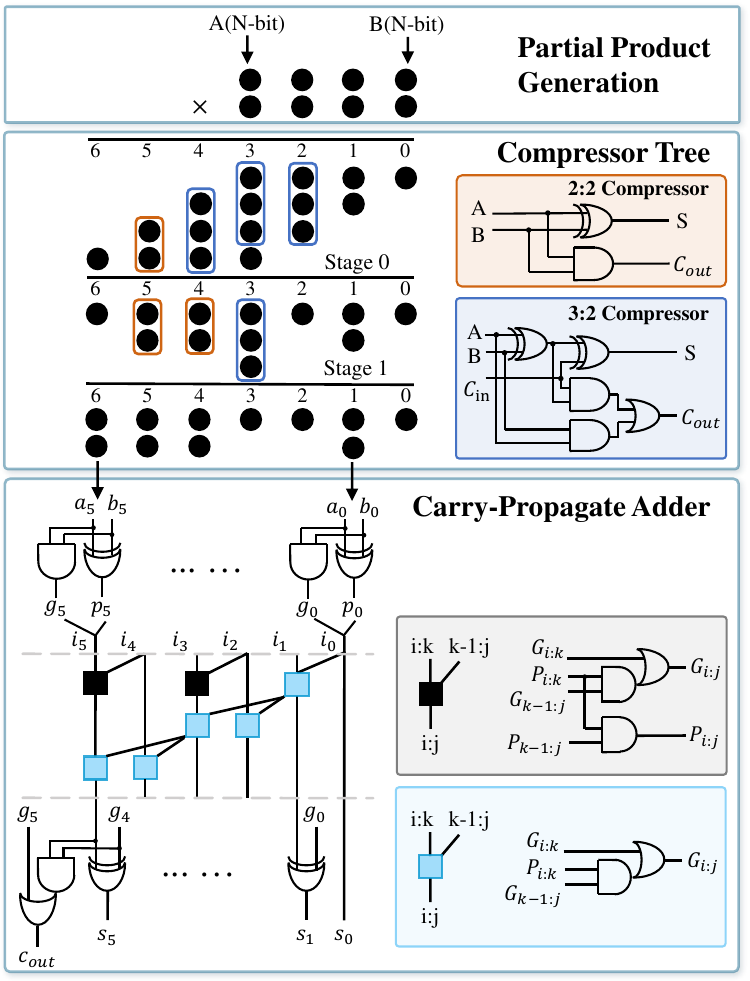}
\vspace{-0.45cm}
\caption{Multiplier Architecture}
\label{fig:multiplier-arch}
\end{figure}

The multiplier architecture integrates three fundamental components: a partial product generator (PPG), a compressor tree (CT) and a carry propagate adder (CPA), as illustrated in \Cref{fig:multiplier-arch}. 

\textbf{Partial Product Generator (PPG):}
The PPG generates partial products (PPs) from multiplicand and multiplier. 
For an \(N\)-bit multiplier, an \textit{AND} gate-based PPG employs \(N^2\) \textit{AND} gates. 
These gates produce PPs, which are shifted according to their bit positions to facilitate subsequent addition.

\textbf{Compressor Tree (CT):}
The primary role of the CT involves compressing the shifted partial products into two parallel rows for parallel reduction. 
It incorporates multiple compression stages, predominantly utilizing 3:2 and 2:2 compressors, which are effectively full adders and half adders, respectively.
A 3:2 compressor at stage \(i\), column \(j\) takes three inputs and outputs a sum to column \(j\) and a carry-out to column \(j+1\) in the next stage \(i+1\). Similarly, a 2:2 compressor at the same stage and column processes two inputs, delivering a sum and a carry-out to the subsequent column and stage.

\textbf{Carry Propagate Adder (CPA):}
The CPA aggregates the two rows of compressed PPs from the CT to produce the final product. 
It generally employs a prefix adder for fast computation.


\subsection{Prefix Structure-based CPA}
\label{sec:prefix_cpa}
The generate function (\(g_i\)) and propagate (\(p_i\)) functions are used in the prefix adders. 
The generate function is the \textit{AND} operation, and the propagate function is the \textit{XOR} operation of the input bits, defined as:

\vspace{-0.55cm}
\begin{equation}
    g_i = a_i \cdot b_i, \quad     p_i = a_i \oplus b_i.
\end{equation}
\vspace{-0.50cm}



The $pg$ functions can be extended to multiple bits and $P_{[i:j]}$, $G_{[i:j]}$ $(i \geq j)$ are defined as:

\vspace{-0.50cm}
\begin{equation}
P_{[i:j]} = 
\begin{cases} 
p_i & \text{if } i = j, \\
P_{[i:k]} \cdot P_{[k-1:j]} & \text{otherwise},
\end{cases}
\end{equation}
\vspace{-0.50cm}

\begin{equation}
G_{[i:j]} = 
\begin{cases} 
g_i & \text{if } i = j, \\
G_{[i:k]} + P_{[i:k]} \cdot G_{[k-1:j]} & \text{otherwise},
\end{cases}
\end{equation}
\vspace{-0.10cm}

The associative operation for the group generate and propagate (\(G, P\)) is defined using the operator \(\circ\):
\vspace{-0.1cm}
\begin{equation}
(G, P)_{[i:j]} = (G, P)_{[i:k]} \circ (G, P)_{[k-1:j]}.
\end{equation}
\vspace{-0.30cm}
The computation of the sum and carry signals is given by:

\begin{equation}
s_i = p_i \oplus c_{i-1}, \quad  c_i = G_{[i:0]} + P_{[i:0]} \cdot c_{in}.
\end{equation}


\subsection{Fused MAC Architecture}
\label{subsec:prelim-fused-mac}
\begin{figure}[!tb]
\centering
\includegraphics[width=.86\linewidth]{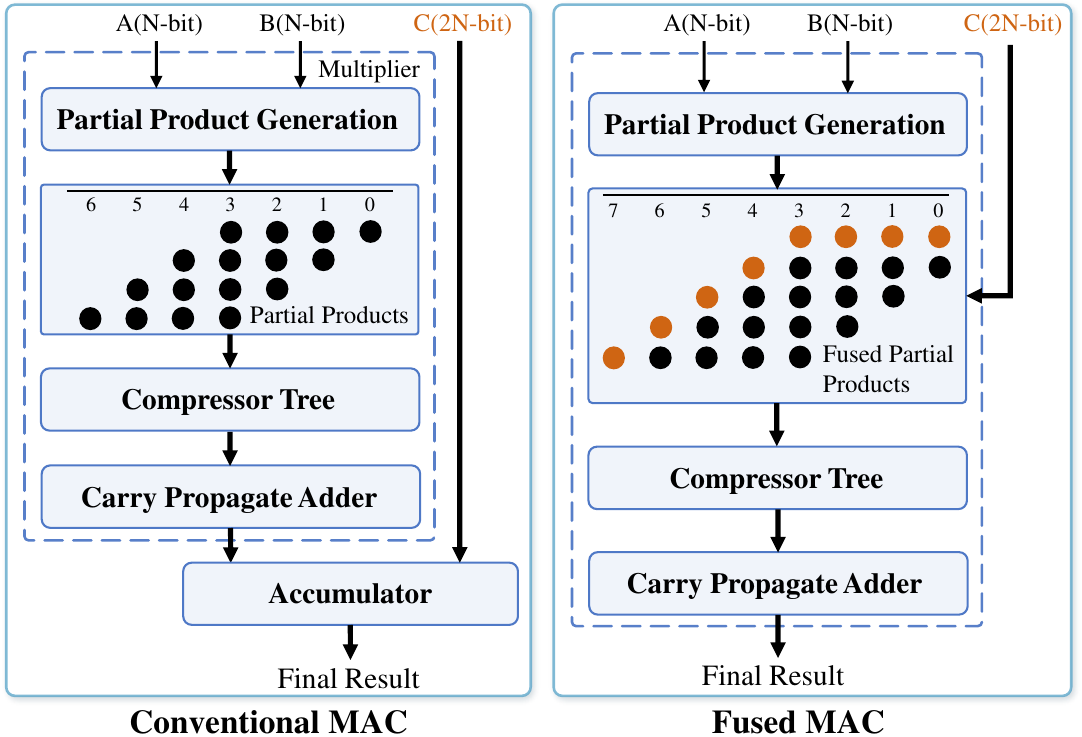}
\vspace{-0.4cm}
\caption{Fused MAC Architecture}
\label{fig:fused-mac-arch}
\end{figure}

As illustrated in \Cref{fig:fused-mac-arch}, the fused multiply-accumulator (fused MAC) architecture integrates the accumulation directly into the compressor tree, eliminating the separate adder stage typically found in conventional MAC units. 
The fusion of the accumulator significantly enhances both area efficiency and delay. 
In this work, we employ the fused MAC architecture to demonstrate its advantages in reducing critical path delay and area, which improves overall efficiency.

\section{Optimization of Compressor Tree}
\subsection{Two Compression Problems}
The CT outputs two rows of compressed partial products, which are fed into CPA to calculate final product results.
So each bit column should output 1 or 2 PPs after compression.
In column $j$, the total number needed to compress is the initial PPs and the carries from column $j-1$ and then compress them to 1 or 2 PPs to produce the final 2 rows of PPs.
This requirement frames our objective in CT optimization: to add the PPs into two rows with minimal cost, a challenge formally described as the Two Compression Problems.

\textbf{Problem:}
    Given an array of initial partial product counts in $2N-1$ columns, denote the number of partial products in column $j$ as $PP_j$. 
    The task is to compress $PP_j + C_{j-1}$ (where $C_{j-1}$ represents the carries from column $j-1$) into a maximum of two outputs per column with minimum total cost. 

In the UFO-MAC framework, we initially determine the optimal counts of the 3:2 and 2:2 compressors for each column. 
We then assign these compressors to stages using ILP, and optimize the interconnection orders between compressors to improve critical path delay. 
These steps are detailed in \Cref{subsec:ct-structure}, \Cref{subsec:ct-assign}, and \Cref{subsec:ct-interconnect}.

\subsection{CT Structure Generation}
\label{subsec:ct-structure}

As described in \Cref{subsec:mult-arch}, 
a 3:2 compressor generates one sum in the current column and passes 1 PP (carry) to the next, which reduces the total number of PPs. 
While a 2:2 compressor is not as efficient as a 3:2 compressor in terms of reducing the total number of PPs. 
For instance, in column $j$, to complete the compression of one PP only by 2:2 compressors, we need to pass it to column $2N$ and require $2N-j$ 2:2 compressors.
Therefore, we use as few 2:2 compressors as possible for more efficient compression.
It is evident that compressing PPs to a single bit incurs higher costs compared to two bits, as more compression requires additional compressors. 
It is ideal to use only 3:2 compressors in columns with even PP numbers ($PP_j + C_{j-1}$ is even), since each 3: 2 compressor reduces 2 PPs in the current column.
However, in columns with odd values $PP_j + C_{j-1}$, it is not feasible to achieve a final count of two using only 3:2 compressors due to parity constraints. 
To adjust parity, we use 2:2 compressors in columns with odd PP numbers~\cite{Datapath-1996TC-Oklobdzija}.
We summarize our CT generation process for each column $j$ in \Cref{alg:ct}.
\begin{algorithm}[!tb]
\caption{Compressor Tree (CT) Generation}\label{alg:ct}
\begin{algorithmic}[1]
\State \textbf{Input:} $PP_j$ for each column $j$, where $j = 0$ to $2N-1$
\State \textbf{Output:} $F_j$ and $H_j$, the counts of 3:2 and 2:2 comps per column
\State Initialize $F_j = 0$ and $H_j = 0$ for all $j$
\For{$j = 0$ to $2N-1$}
    \If{$j = 0$}
        \State Adjust $C_{-1} = 0$ \Comment{Initial carry for the first column}
    \EndIf
    \If{$(PP_j + C_{j-1})$ is even} \label{alg:ct:even}
        \State $F_j \leftarrow (PP_j + C_{j-1} - 2) / 2$
    \Else \Comment{Odd number of PPs}
        \State $H_j \leftarrow 1$ \Comment{Adjust for parity}
        \State $F_j \leftarrow (PP_j + C_{j-1} - 3) / 2$
    \EndIf
\EndFor
\end{algorithmic}
\end{algorithm}


The gate-level structures of the 3:2 and 2:2 compressors are illustrated in \Cref{fig:multiplier-arch}. 
In CMOS technology, the \textit{AND} and \textit{OR} logic is typically implemented with \textit{NAND} and \textit{OAI} gates. 
Thus, the area of a 3:2 compressor is typically 1.5 times that of a 2:2 compressor. 
For a column with $M$ bit total PPs, the minimum number of compression stages required is given by $\left\lceil \log_{\frac{3}{2}} \left(\frac{M}{2}\right) \right\rceil$~\cite{Datapath-1964TC-Wallace}. 
As described above, we only allow for no more than one 2:2 compressor in each column.
Considering both area and stage requirements and the 2:2 compressor number constraints, we next demonstrate that our CT design is optimal, minimizing both the area and the number of stages.
First, we prove that our approach has a minimum CT area.

%

\begin{proof}
Let $F$ and $H$ be the numbers of 3:2 and 2:2 compressors, respectively, in our supposed optimal design with an area of $3F + 2H$. 
Assume that there exists a compressor tree that uses fewer 3:2 or 2:2 compressors and still meets the two-output maximum per column. 
Removing $m$ 3:2 compressors would result in $2+2m$ outputs in the affected columns, exceeding the limit of two outputs per column and thus violating the constraints of the problem. 
Similarly, removing a single 2:2 compressor from the columns where exactly one is used would leave $3$ outputs ($2+1$), again violating the constraints of the problem.
Substituting one 2:2 compressor with one 3:2 compressor would result in an area of $3F + 2H + 1$, thereby increasing the total area.
Replacing $x$ 3:2 compressors with $y$ 2:2 compressors, where $2x \leq y$, results in an area change of $3F + 2H - 3x + 2y$. 
This increases the area since $3F + 2H - 3x + 2y \geq 3F + 2H + x$, thus proving by contradiction that our original design is optimal by minimizing the compressor area without violating any design constraints.
\end{proof}
\vspace{-0.2cm}
Next, we prove that our approach has a minimum stage number:
\begin{proof}[Proof of Minimum Compressors per Column]
For any given column $j$ in a compressor tree, let $pp_j + C_{j-1}$ be the total number of partial products and carries to be compressed. 
Assume our solution, which uses $F_j$ 3:2 compressors and $H_j$ 2:2 compressors, and suppose that there is a feasible solution with fewer compressors. 
Reducing any 3:2 compressor by $m$ would result in excess outputs (more than two). 
Similarly, reducing a 2:2 comp, since $h_j \leq 1$, would result in more than two outputs for that column, violating the two output constraint. 
Adjusting the compressor configuration by replacing $m$ 3:2 comps with $n$ 2:2 compressors to maintain constraints would require $n = 2m$. 
This replacement results in a compressor count of $f_j + h_j + m$, which is greater than the original count, proving by contradiction that our compressor allocation for each column is minimal.
\end{proof}
\vspace{-0.5cm}
\begin{proof}[Proof of Minimum Stages in the Compressor Tree]
Having established that each column is compressed using the minimum number of compressors, it follows that the carry propagated to the next column is also minimized. 
Each additional compressor could potentially introduce an additional stage of the next column due to propagation of its carry. Since our arrangement of compressors is minimal for each column, and no unnecessary carries are generated, the entire tree achieves a minimal stage count. The number of stages required can be calculated using the formula $\left\lceil \log_{\frac{3}{2}}\left(\frac{N}{2}\right)\right\rceil$, where $N$ combines $pp_j$ and $C_{j-1}$.
\end{proof}
\vspace{-0.2cm}
Previous work such as GOMIL~\cite{Datapath-2021DATE-Xiao} utilizes the one- or two-bit output from the CT to reduce the need for $pg$ generation logic in the CPA. 
However, the reduction in $pg$ logic leads to an additional 3:2 compressor in the CT, which does not result in overall area savings. 
This is because the $pg$ generation logic is typically implemented using one NOR and two NAND gates and occupies less area than a 3:2 compressor.

\subsection{Compressor Assignment}
\label{subsec:ct-assign}

Building on the optimal counts of the 3:2 and 2:2 compressors for each column by \Cref{alg:ct}, we introduce a method to assign these compressors to specific stages, thus achieving a compressor tree structure with a minimized stage count. 
Previous efforts such as GOMIL~\cite{Datapath-2021DATE-Xiao} do not account for the number of stages, and heuristic assignments in RL-MUL~\cite{Datapath-2023DAC-Zuo} potentially result in suboptimal stage utilization. 
In contrast, our approach employs an ILP model to determine the stage assignments that minimize the total number of CT stages.

We define $Slice_{i,j}$ as the set of compressors located at stage $i$ and column $j$ in the compressor tree. 
And we set a stage limit, $stage_max$. 
For each $Slice_{i,j}$, the assigned numbers of 3:2 and 2:2 compressors are represented by $f_{i,j}$ and $h_{i,j}$, respectively. 
We ensure that the total compressors across all stages match the given counts from \Cref{alg:ct} with the following constraints:
\begin{equation}
    \sum_{i=0}^{\text{stage\_max}} f_{i,j} = F_j \quad \forall j
\end{equation}
\begin{equation}
    \sum_{i=0}^{\text{stage\_max}} h_{i,j} = H_j \quad \forall j
\end{equation}
We define $pp_{i,j}$ as the number of PPs at $Slice_{i,j}$. 
The PPs within each slice are influenced by the outputs from the previous stage and the carries from the preceding column, leading to the constraint:
\begin{equation}
pp_{i,j} = pp_{i-1,j} - 2f_{i,j} - h_{i,j} + f_{i-1,j-1} + h_{i-1,j-1}, \quad \forall i > 0, \forall j > 0
\end{equation}
Furthermore, the number of PPs must be sufficient to accommodate the compressors within a slice:
\begin{equation}
3f_{i,j} + 2h_{i,j} \leq pp_{i,j}, \quad \forall i, j
\end{equation}
To minimize the total number of stages, $S$, we use a sufficiently large constant $M$ and binary auxiliary variables $y_{i,j}$ to indicate whether any compressor is placed at $Slice_{i,j}$:
\begin{equation}
S \geq i \cdot y_{i,j}, \quad \forall i, j
\end{equation}
\begin{equation}
M\cdot y_{i,j} \geq f_{i,j} + h_{i,j}, \quad \forall i, j
\end{equation}
Our primary objective is to reduce the number of stages in the compressor tree:
\begin{equation}
\min \quad S
\end{equation}
By incorporating boundary conditions, this formulation allows for deriving the CT structure with the minimum number of stages.

\subsection{Impact of Interconnection Order}
\begin{figure}[!tb]
\centering
\includegraphics[width=.72\linewidth]{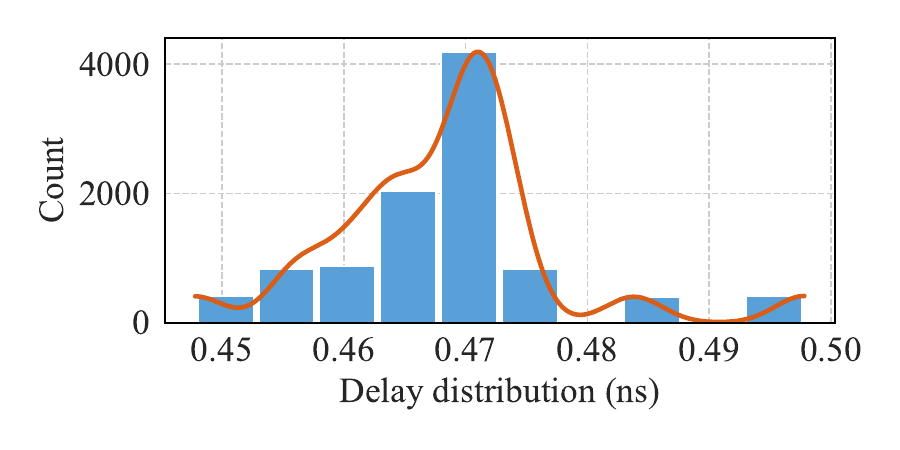}
\caption{Critical path delay distribution of 10000 random interconnect order with one same CT stage structure.}
\label{fig:interconnect-distr}
\end{figure}

The interconnection order between compressors can affect the critical path delay of the CT, which represents a design space that previous works have often overlooked.
As illustrated in \Cref{fig:multiplier-arch}, for a 3:2 compressor, the path from ports $A$ and $B$ to port $Sum$ involves two \textit{XOR} gates, whereas the path from $C_{in}$ to $C_{out}$ passes through \textit{AND} and \textit{OR} logic, implemented by \textit{NAND} and \textit{OAI} gates. 
In particular, the delay through two \textit{XOR} gates is approximately 1.5 times that of the \textit{NAND} and \textit{OAI} combination. 
Furthermore, the delay of 2:2 compressors is less than that of 3:2 compressors since they only pass through one \textit{XOR} or one \textit{AND} gate. 
To demonstrate the impact of interconnection order, we assign 10,000 random interconnection orders to the same CT structure, and then synthesize the 10,000 CTs with the same constraints.
As shown in \Cref{fig:interconnect-distr}, the synthesized results indicated that the delay of the CT varied by over 10\%.

\begin{figure*}[!tb]
\centering
\includegraphics[width=0.95\linewidth]{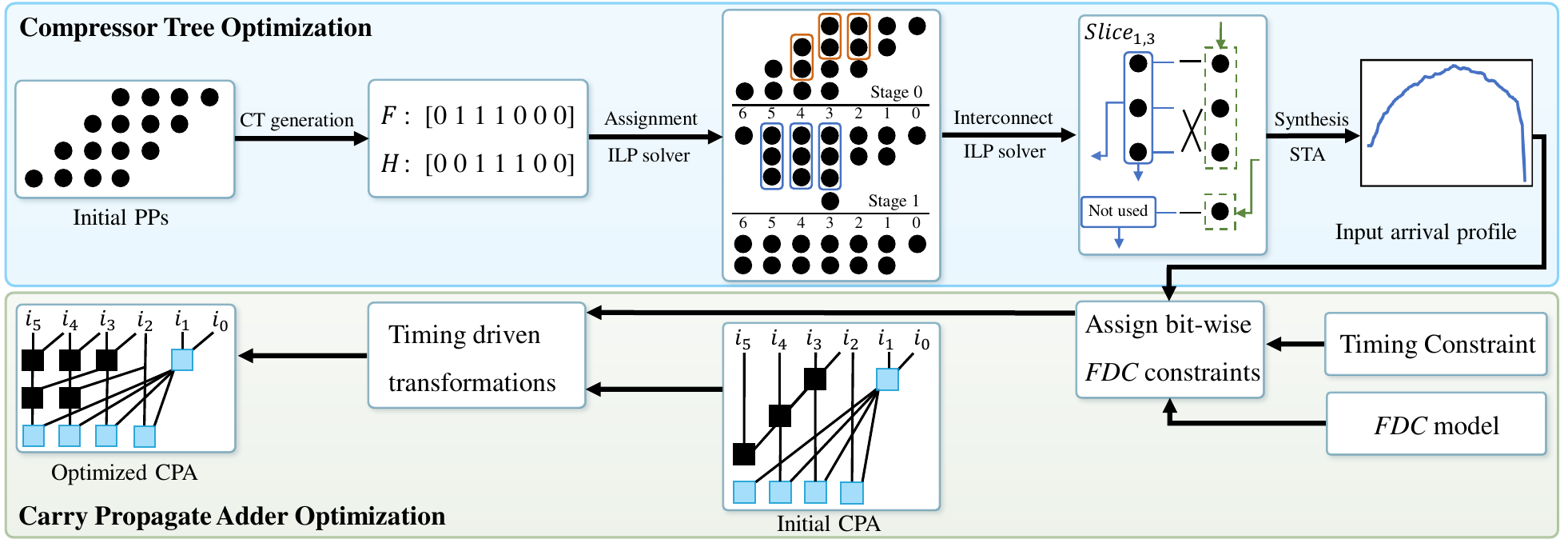}
\vspace{-0.42cm}
\caption{UFO-MAC framework. The framework first generates optimal CT structures and then performs timing-driven optimizations on the CPA based on a non-uniform arrival profile to achieve area-delay efficiency.} 
\label{fig:framework}
\end{figure*}

\subsection{Interconnection Order Optimization}
\label{subsec:ct-interconnect}
We propose an ILP-based approach to optimize the interconnection orders of compressors within the compressor tree. 
Considering a 3:2 compressor, we assume that the arrival times of the input at ports $A$, $B$, and $C_{in}$ are $a$, $b$, and $d$ respectively. 
The output timing for the sum and carry can then be determined by the following equations:
\begin{equation}
s = \max(a + T_{as}, b + T_{bs}, d + T_{cs})
\label{equ:s_const}
\end{equation}
\begin{equation}
c = \max(a + T_{ac}, b + T_{bc}, d + T_{cc})
\label{equ:c_const}
\end{equation}
Here, $T_{xy}$ represents the delay from input $x$ to output $y$.
We transform the maximum operations into linear constraints:
\begin{equation}
s \geq a + T_{as},\quad s \geq b + T_{bs},\quad s \geq d + T_{cs}
\label{equ:s_const_linear}
\end{equation}
\begin{equation}
c \geq a + T_{ac},\quad c \geq b + T_{bc},\quad c \geq d + T_{cc}
\label{equ:2}
\end{equation}
Similarly, these constraints are applicable to 2:2 compressors, with corresponding adjustments for their specific input and output timing characteristics. 

\begin{figure}[!tb]
\centering
\includegraphics[width=.92\linewidth]{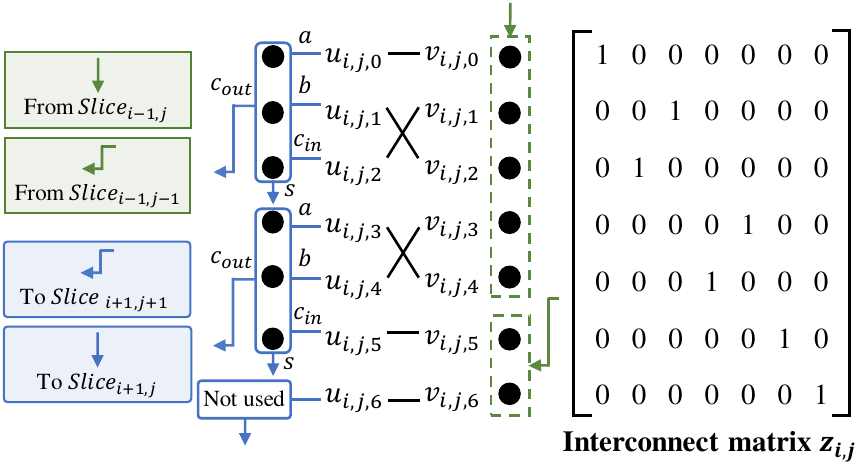}
\vspace{-0.5cm}
\caption{Interconnection order slice example}
\label{fig:interconnect}
\end{figure}

As illustrated in \Cref{fig:interconnect}, the PPs in $Slice_{i,j}$ originate from two sources: the sums and unused PPs from $Slice_{i-1,j}$, and the carries from $Slice_{i-1,j-1}$.
For $Slice_{i,j}$, which receives $m$ PPs in total, we denote the PPs of $Slice_{i,j}$ as a source vector $u_{i,j}$, representing the data arrival times:
\begin{equation}
    [pp_{i,j,0},\quad pp_{i,j,1},\quad \ldots, pp_{i,j,m-1}]
\end{equation}
Consequently, PPs can connect to compressor ports or directly pass to $Slice_{i+1,j}$. 
We assign dummy ports for these PPs, which are not used and left to $Slice_{i+1,j}$. 
The arrival times of these connections are denoted in the sink vector $v_{i,j}$:
\begin{equation}
    [port_{i,j,0},\quad port_{i,j,1},\quad \ldots, port_{i,j,m-1}]
\end{equation}
The task is to optimize the bijective mapping between the source vector $u_{i,j}$ and the sink vector $v_{i,j}$ for each $Slice_{i,j}$. 
To model the bijection between the source and sink vectors in each slice, we introduce a $m \times m$ binary matrix $z_{i,j}$. 
Each entry $z_{i,j,u,v} = 1$ indicates that the source $u$ is connected to the sink $v$. 
The formulation of this relationship is given by:
\begin{equation}
    v = u \quad \text{if and only if} \quad z_{i,j,u,v} = 1
    \label{z_constr0}
\end{equation}
To transform the constraints in \Cref{z_constr0} linearly, we employ a sufficiently large constant $Z$, and linear constraints are as follows:
\begin{equation}
    v - u \leq Z \cdot (1 - z_{i,j,u,v}); \quad     u - v \leq Z \cdot (1 - z_{i,j,u,v})
    \label{equ:big_M_linear}
\end{equation}
The following constraints ensure that each input is connected to exactly one output and vice versa:
\begin{equation}
    \sum_{v=0}^{m-1} z_{i,j,u,v} = 1, \quad \forall u;  \quad
    \sum_{u=0}^{m-1} z_{i,j,u,v} = 1, \quad \forall v
    \label{equ:z_constr2}
\end{equation}

Then combined with \Cref{equ:s_const_linear,equ:2}, we can get the data arrival time of every partial product and every compressor port.
To minimize the critical path in the compressor tree for multipliers of $N$ bits, the objective is to reduce the longest delay among the final outputs. 
We define $M$ as the maximum delay in any of the columns from $0$ to $2N-1$. The goal is formulated as minimizing this maximum delay, represented mathematically by:
\begin{equation}
    M \geq t_{j,0}, \quad M \geq t_{j,1} \quad \text{for all} \quad j \in \{0, 2N-1\}
\end{equation}
\begin{equation}
    \min \quad M
\end{equation}
The ILP formulation can handle all initial partial product shapes, and we can easily extend to optimization of CT of fused MAC.

\section{Optimization of CPA}

Building upon the optimized compressor tree structures, we have developed a refined approach for CPA design. 
This method effectively utilizes the non-uniform arrival profile of the CPA to achieve area-delay efficiency. 
Our comprehensive framework, as shown in \Cref{fig:framework}, integrates these optimizations into the design process. 
This section will detail our methods for CPA optimization.

\subsection{Non Uniform Arrival Profile of CPA}
As illustrated in \Cref{fig:motivation}, the carry propagation adder presents a non-uniform arrival profile, presenting unique challenges in design and optimization compared to CPAs with uniform profiles. 
Leveraging the variance in data arrival times, we aim to create area-delay efficient adders that conform to timing constraints.
Our refined optimization framework explicitly exploits non-uniform arrival times. 
Initially, area-efficient adder structures are selected, followed by timing-driven transformations to meet the constraints.

The CPA's arrival profile is segmented into three regions as shown in \Cref{fig:motivation}:

\textbf{Region 1:} With a ``positive slope'', where faster adders are unnecessary, we employ a Ripple Carry Adder (RCA) suitable for gradual arrival times.

\textbf{Region 2:} Known as the flat region with the latest data arrivals, necessitating fast adder structures like the Sklansky structure \cite{Datapath-1960TC-Sklansky}.

\textbf{Region 3:} Characterized by a ``negative slope'' in which data at the MSB end arriving first. 
To align with this ``negative slope'', we use a Carry Increment Adder \cite{1996-IWLAS-Zimmermann} as the initial structure.

The initial structure effectively utilizes the non-uniform profile to optimize both area and delay across different regions of the CPA.

\subsection{Timing Modeling for Prefix Adders}
\label{subsec:timing-adder}
\begin{figure}[!tb]
\begin{minipage}{.52\linewidth}
    \centering
    \includegraphics[width=.98\linewidth]{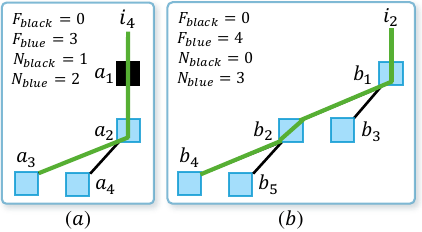}
    \vspace{-0.45cm}
    \captionof{figure}{Sub-prefix trees.}
    \label{fig:mpfo}
\end{minipage}
\begin{minipage}{.46\linewidth}
    \centering
    \begin{tabular}{ccc}
    \toprule
    Feature & $R^2$ Score & MAPE\\
    \midrule
    logic depth & 0.541 &  9.30\% \\
    mpfo & 0.469 & 10.91\% \\
    \textbf{FDC} & 0.816 &  4.63\% \\
    \bottomrule
    \end{tabular}
    \captionof{figure}{Timing Model.}
    \label{tab:timing_fit}
\end{minipage}
\end{figure}

Following the selection of initial area-efficient structures, we refine them based on timing constraints. 
To ensure that each bit's critical path in the CPA meets timing constraints, we extract a sub-prefix tree from a specific bit position to estimate and optimize the critical path delay for that bit.
\Cref{fig:mpfo} shows trees extracted from bit positions 1 and 3 of the CPA in \Cref{fig:multiplier-arch}.
Once the prefix tree is extracted, we can estimate the delay for further optimizations. 
High-fidelity timing modeling is crucial to achieving accurate delay estimations. 
Many previous works have used logic depth as a timing model \cite{1996-IWLAS-Zimmermann,Datapath-2007GLSVLSI-Matsunaga,Datapath-2021DATE-Xiao}. 
The max-path-fanout (mpfo) was introduced in \cite{DSE-2019TCAD-MA}, which accumulates the fanout count of each node along a path, and does not take into account the logic depth.
Recognizing that path delay is influenced by both logic depth and fanout, and that existing models overlook the distinct node types shown in \Cref{fig:multiplier-arch}, we introduce the \textbf{fanout depth combination (FDC)}. This refined model integrates path depth, fanout, and node types to offer a more accurate and comprehensive timing prediction, addressing the limitations of previous models.

We apply the simplified logic effort method\cite{Datapath-2003Asilomar-Harris} for timing estimation as follows:
\begin{equation}
d = g \times f + p
\end{equation}
where \( g \) is the logic effort, \( f \) the fanout, and \( p \) the intrinsic delay of the gate. 
This model is adapted for different types of nodes, we denote $g_{black}$, $ p_{black}$ and $g_{blue}$, $p_{blue}$ as the logic effort and intrinsic delay of black and blue nodes, respectively. 
Black nodes encompass \textit{AND-OR} logic and \textit{AND} logic, implemented through interleaving \textit{AOI+NAND} and \textit{OAI+NOR}.
In contrast, blue nodes are implemented using only \textit{AOI} or \textit{OAI} cells. 
For black nodes, the delay is:
\begin{equation}
d_{black} = g_{black} \times (f_{black} + f_{blue}) + p_{black}
\label{equ:delay_black}
\end{equation}
where \( f_{black} \) and \( f_{blue} \) are the fanouts to black and blue nodes, respectively.
Blue nodes, typically final level nodes only driving a single sum logic, making their delay a constant:
\begin{equation}
d_{blue} = g_{blue} \times f_{sum} + p_{blue}
\label{equ:delay_blue}
\end{equation}
where \( f_{sum} \) is the fanout to sum logic and is set to one. 
By integrating these with \Cref{equ:delay_black,equ:delay_blue}, the critical path delay for a tree starting from bit $i$ can be represented as:
\begin{equation}
d_{i} = k_0 \times F_{black} + k_1 \times F_{blue} + k_2 \times N_{black} + k_3 \times N_{blue} + b
\end{equation}
Here, $k_0, k_1, k_2, k_3,$ and $b$ are coefficients that can be determined to fit the model.
Examples of FDC features are shown in highlighted paths in \Cref{fig:mpfo}.

To determine the maximum depth, mpfo, and FDC in a tree consisting of $n$ nodes, the computational complexity for each method is $O(n)$.
To validate the fidelity of FDC, we conducted linear regression analyses for the depth model, mpfo, and FDC, comparing the $R^2$ Score and Mean Absolute Percentage Error (MAPE). 
These analyzes are based on 10,000 paths extracted from the open-source adder dataset comprising 1100 adders from \cite{DSE-2019TCAD-MA}. 
The results, presented in \Cref{tab:timing_fit}, show that by incorporating fanout and node types, FDC significantly improves fidelity within the same computational complexity.

\subsection{Final Adder Optimization}
\begin{figure}[!tb]
\centering
\includegraphics[width=1\linewidth]{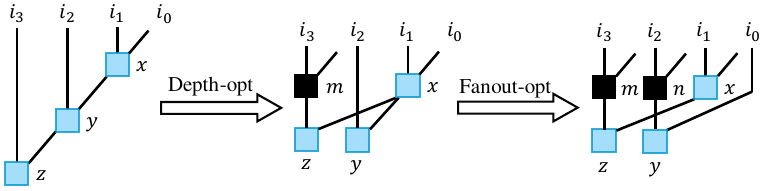}
\vspace{-0.7cm}
\caption{Example of two optimization transformations.}
\label{fig:transformation}
\end{figure}

Based on the optimized CT structure, the non-uniform arrival time of the CPA is normalized to the FDC model scale, and maximum FDC constraints are set for each input bit based on timing requirements (\Cref{subsec:timing-adder}). 
Then iterative timing-driven optimization is applied to meet these constraints, employing the depth-opt and fanout-opt transformations illustrated in \Cref{fig:transformation}.

\begin{algorithm}[!tb]
\caption{Timing-driven Prefix Graph Optimization}
\label{alg:cpa}
\begin{algorithmic}[1]
    \State \textbf{Input:} Input arrival times $A_j$ for each bit $j$, timing constraints $C$, initial prefix graph $G$, FDC timing model \label{alg:cpa:1}
    \State \textbf{Output:} Optimized prefix graph $G'$ \label{alg:cpa:2}
    \State Assign bit-wise FDC constraints $c_j$ for each bit $j$ \label{alg:cpa:3}
    \While{all $c_j$ are met and exist possible optimization} \label{alg:cpa:4}
        \For{$j = MSB$ to $LSB$} \Comment{Iterate from MSB to LSB} \label{alg:cpa:5}
            \If{$c_j$ are violated} \label{alg:cpa:6}
                \State Extract sub-prefix tree $T_j$ starting from bit $j$ \label{alg:cpa:7}
                \If{Depth of $T_j > \log_2(N)$} \Comment{check min depth} \label{alg:cpa:8}
                    \State $p \gets $ node with maximum depth in $T_j$ \label{alg:cpa:9}
                    \State \Call{GraphOpt}{$p$} \Comment{depth-opt} \label{alg:cpa:10}
                \Else \label{alg:cpa:11}
                    \State $p \gets $ node with maximum siblings in $T_j$ \label{alg:cpa:12}
                    \State \Call{GraphOpt}{$p$} \Comment{fanout-opt} \label{alg:cpa:13}
                \EndIf \label{alg:cpa:14}
            \EndIf \label{alg:cpa:15}
        \EndFor \label{alg:cpa:16}
    \EndWhile \label{alg:cpa:17}
    \State \Return $G'$ \Comment{Return the optimized graph} \label{alg:cpa:18}
\Procedure{GraphOpt}{$p$} \label{alg:cpa:19}
    \State Create a new node $s$ \label{alg:cpa:20}
    \State $ntf(s) \leftarrow tf(ntf(p))$, $ntf(s) \leftarrow tf(ntf(p))$ \label{alg:cpa:21}
    \State $tf(p) \leftarrow s$, $ntf(p) \leftarrow ntf(ntf(p))$ \label{alg:cpa:22}
\EndProcedure \label{alg:cpa:23}
\end{algorithmic}
\end{algorithm}

Each prefix node $p$ has two fan-ins: the trivial fan-in ($tf$), which is vertically aligned and shares the same MSB, and the non-trivial fan-in ($ntf$). 
We denote $tf(y)$ and $ntf(y)$ as trivial and non-trivial fan-ins of $y$.
For example, in the prefix graph on the left side of \Cref{fig:transformation}, $tf(y)$ and $ntf(y)$ refer to $i_2$ and $x$, respectively.

Recognizing the influence of logic depth and fanout on path delay, we propose two optimization strategies: \textbf{depth optimization} (depth-opt) and \textbf{fanout optimization} (fanout-opt).
While prior refine-based works \cite{Datapath-1990DAC-Fishburn, 1996-IWLAS-Zimmermann} primarily focused on depth, the significance of fanout optimization has often been neglected. 
Our approach addresses this oversight by balancing both aspects, effectively managing the trade-offs between logic depth, node count, and fanout for improved timing and area efficiency \cite{Datapath-2003Asilomar-Harris}.
The specific rules for implementing these transformations are detailed in \Cref{alg:cpa:19,alg:cpa:20,alg:cpa:21,alg:cpa:22,alg:cpa:23}, with the same principles applying to both depth-opt and fanout-opt. The key distinction lies in the nodes targeted for optimization.

Our timing-driven prefix graph optimization strategy is described in \Cref{alg:cpa}. 
The algorithm adjusts the prefix graph from the MSB to the LSB to resolve timing violations(\Cref{alg:cpa:4}).
The algorithm checks each bit for timing constraints, and bits with timing violations, it extracts the sub-prefix tree from the bit (\Cref{alg:cpa:6,alg:cpa:7}). 
For a prefix tree that spans $N$ bits, the minimal depth is given by $\log_2(N)$ \cite{Datapath-1986JA-Snir}.
Depending on the depth of the tree, the optimization method is chosen: If the tree depth exceeds $\log_2(N) +1$ (plus 1 for nodes to group $PG$ from the LSB side), indicating depth inefficiency, depth optimization is applied to reduce depth.
Otherwise, if the depth is already optimal or minimal, fanout optimization is performed to balance high fanout nodes in the tree (\Cref{alg:cpa:11,alg:cpa:12,alg:cpa:13}).
This process continues iteratively until all bits meet the timing constraints or no further optimizations are possible, ensuring that the prefix graph is optimized for both area and delay.
\section{Experimental Results}
\subsection{Setup}
\begin{figure}[!tb]
\centering
\includegraphics[width=0.816\linewidth]{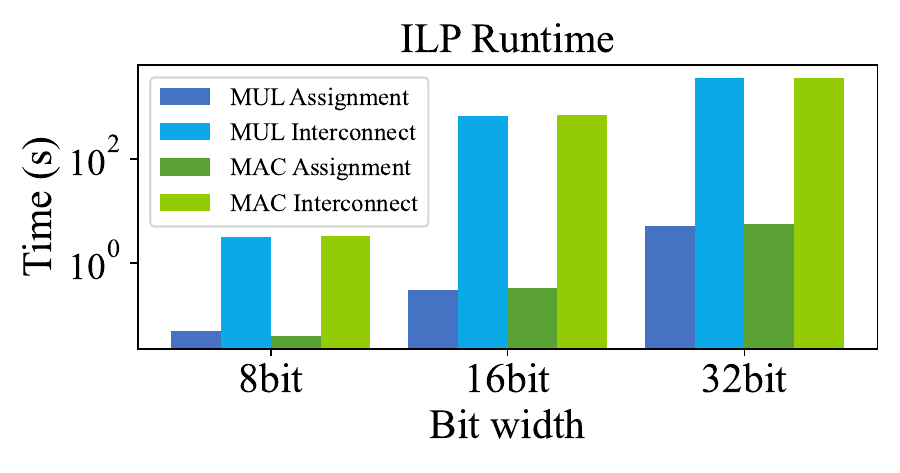}
\caption{ILP runtime.}
\label{fig:runtime}
\end{figure}

The proposed framework is implemented on a Linux platform with a 2.0GHz Intel Xeon Gold 6338 CPU with 1024GB of memory and an NVIDIA RTX 4090 GPU. 
The obtained designs are functionally correct which is verified by equivalence checking in Berkerly ABC~\cite{Berkerly-ABC}.
For ILP solvers, we use the Gurobi Optimizer (version 11.0) \cite{Gurobi} and set the ILP runtime limit to 3,600 seconds with 128 threads for compressor assignment and interconnect order optimizations, and the detailed runtime is shown in \Cref{fig:runtime}.
For each bitwidth configuration of multipliers and MACs, we use timing-driven, area-driven, and trade-off strategies for CPA optimization in \Cref{alg:cpa}.
Comparisons are drawn between compressor trees, multipliers, and MACs generated by UFO-MAC and baseline approaches. 
Our baselines include:

\textbf{GOMIL}\cite{Datapath-2021DATE-Xiao}: An ILP-based global optimization method. Given GOMIL's special prefix node implementation, we execute ILP and generate RTL code using the provided open-source C++ code and set the ILP runtime to 10,000 seconds with 128 threads.

\textbf{RL-MUL}\cite{Datapath-2023DAC-Zuo}: A state-of-the-art RL-based approach. We reproduce the RL framework, running it for 3,000 steps as specified in RL-MUL. Given its focus solely on CT optimization, we follow the original setting to use default adders from synthesis tools.


\textbf{Commercial IP}: We utilize $y=a*b$ and $y=a*b+c$ style RTL and commercial tools/IPs for synthesis. For compressor tree comparisons, we instantiated commercial compressor tree IP in the RTL.

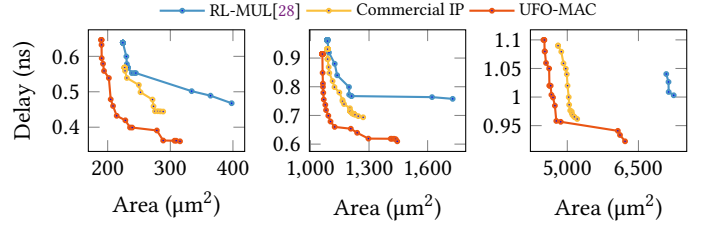
\begin{figure}[!tb]
    \centering

\definecolor{CUHKorange}{RGB}{244,106,18} 
\definecolor{CUHKblue}{RGB}{0,111,190}    
\definecolor{CUHKgreen}{RGB}{0,127,128}   
\definecolor{CUHKmiddle}{RGB}{144,44,144} 
\definecolor{CUHKdark}{RGB}{114,44,114}   
\definecolor{CUHKred}{RGB}{228,46,36}     
\definecolor{CUHKyellow}{RGB}{198,148,34} 
\definecolor{myblue}{RGB}{73,148,196}   
\definecolor{mydarkblue}{RGB}{18,38,79} 
\definecolor{myorange}{RGB}{234,85,20}  
\definecolor{myyellow}{RGB}{250,192,61} 
\definecolor{mypink}{RGB}{252,228,215}  
\definecolor{mygreen}{RGB}{19,138,7}  

\begin{tikzpicture}
\begin{groupplot}[group style={group size= 3 by 1, horizontal sep=0.86cm, group name=myplot}, height=3.2cm, width=3.66cm]
\nextgroupplot[minor tick num=0,
xlabel={Area ($\upmu$m$^2$)},
ylabel={Delay (ns)},
y label style={at={(-0.2,0.5)}},
ylabel near ticks,
legend style={
    draw=none,
	at={(0.5,1.15)},
	nodes={scale=0.75, transform shape},
	anchor=north,
	legend columns=-1,
}
]   
    \addplot[blue, style={mark=*, mark size=0.7pt, line width=0.8pt, draw=myblue}]  table [x=Cell Area, y=Critical Path Slack, col sep=comma] {pgfplot/data/ct/8bit/rl_pareto_output.csv};
    \addplot[blue, style={mark=*, mark size=0.7pt, line width=0.8pt, draw=myyellow}]  table [x=Cell Area, y=Critical Path Slack, col sep=comma] {pgfplot/data/ct/8bit/dw_pareto_output.csv};
    \addplot[blue, style={mark=*, mark size=0.7pt, line width=0.8pt, draw=myorange}]  table [x=Cell Area, y=Critical Path Slack, col sep=comma] {pgfplot/data/ct/8bit/Ours_pareto_output.csv};
\coordinate (left) at (rel axis cs:0,1);

\nextgroupplot[minor tick num=0,
xlabel={Area ($\upmu$m$^2$)},
xtick={1000,1300,1600},
y label style={at={(-0.2,0.5)}},
ylabel near ticks,
legend style={
    draw=none,
	at={(0.45,1.3)},
	nodes={scale=0.75, transform shape},
	anchor=north,
	legend columns=-1,
}
]   
    \addplot[blue, style={mark=*, mark size=0.7pt, line width=0.8pt, draw=myblue}]  table [x=Cell Area, y=Critical Path Slack, col sep=comma] {pgfplot/data/ct/16bit/rl_pareto_output.csv};\addlegendentry{RL-MUL\cite{Datapath-2023DAC-Zuo}};
    \addplot[blue, style={mark=*, mark size=0.7pt, line width=0.8pt, draw=myyellow}]  table [x=Cell Area, y=Critical Path Slack, col sep=comma] {pgfplot/data/ct/16bit/dw_pareto_output.csv};\addlegendentry{Commercial IP};
    \addplot[blue, style={mark=*, mark size=0.7pt, line width=0.8pt, draw=myorange}]  table [x=Cell Area, y=Critical Path Slack, col sep=comma] {pgfplot/data/ct/16bit/ours_pareto_output.csv};\addlegendentry{UFO-MAC};
\coordinate (mid) at (rel axis cs:0.5,1);

\nextgroupplot[minor tick num=0,
xlabel={Area ($\upmu$m$^2$)},
xtick={5000,6500},
y label style={at={(-0.2,0.5)}},
ylabel near ticks,
legend style={
    draw=none,
	at={(0.9,1.2)},
	nodes={scale=0.75, transform shape},
	anchor=north,
	legend columns=-1,
}
]
    \addplot[blue, style={mark=*, mark size=0.7pt, line width=0.8pt, draw=myblue}]  table [x=Cell Area, y=Critical Path Slack, col sep=comma] {pgfplot/data/ct/32bit/rl_pareto_output.csv};
    \addplot[blue, style={mark=*, mark size=0.7pt, line width=0.8pt, draw=myyellow}]  table [x=Cell Area, y=Critical Path Slack, col sep=comma] {pgfplot/data/ct/32bit/dw_pareto_output.csv};
    \addplot[blue, style={mark=*, mark size=0.7pt, line width=0.8pt, draw=myorange}]  table [x=Cell Area, y=Critical Path Slack, col sep=comma] {pgfplot/data/ct/32bit/ours_pareto_output.csv};
\coordinate (right) at (rel axis cs:1,1);

\end{groupplot}
\path (left)--(right) coordinate[midway] (group center);
\path (myplot c2r1.north west|-current bounding box.north)--
coordinate(legendpos)
(myplot c2r1.north west|-current bounding box.north);
\end{tikzpicture}
    \vspace{-0.8cm}
    \caption{Pareto-frontiers of the synthesized results on compressor trees. From left to right: 8-bit; 16-bit; 32-bit.}
    \label{fig:results-ct}
\end{figure}

\begin{figure}[!tb]
    \centering

\definecolor{CUHKorange}{RGB}{244,106,18} 
\definecolor{CUHKblue}{RGB}{0,111,190}    
\definecolor{CUHKgreen}{RGB}{0,127,128}   
\definecolor{CUHKmiddle}{RGB}{144,44,144} 
\definecolor{CUHKdark}{RGB}{114,44,114}   
\definecolor{CUHKred}{RGB}{228,46,36}     
\definecolor{CUHKyellow}{RGB}{198,148,34} 

\definecolor{myblue}{RGB}{73,148,196}   
\definecolor{mydarkblue}{RGB}{18,38,79} 
\definecolor{myorange}{RGB}{234,85,20}  
\definecolor{myyellow}{RGB}{250,192,61} 
\definecolor{mypink}{RGB}{252,228,215}  
\definecolor{mygreen}{RGB}{19,138,7}  


\begin{tikzpicture}
\begin{groupplot}[group style={group size= 3 by 1, horizontal sep=0.86cm, group name=myplot}, height=3.2cm, width=3.66cm]
\nextgroupplot[minor tick num=0,
xlabel={Area ($\upmu$m$^2$)},
ylabel={Delay (ns)},
y label style={at={(-0.2,0.5)}},
ylabel near ticks,
legend style={
    draw=none,
	at={(0.25,1.)},
	nodes={scale=0.75, transform shape},
	anchor=north,
	legend columns=-1,
}
]   
    \addplot[blue, style={mark=*, mark size=0.7pt, line width=0.8pt, draw=mydarkblue}]  table [x=Cell Area, y=Critical Path Slack, col sep=comma] {pgfplot/data/mul/8bit/gomil_pareto_output.csv};
    \addplot[blue, style={mark=*, mark size=0.7pt, line width=0.8pt, draw=myblue}]  table [x=Cell Area, y=Critical Path Slack, col sep=comma] {pgfplot/data/mul/8bit/rl_pareto_output.csv};
    \addplot[blue, style={mark=*, mark size=0.7pt, line width=0.8pt, draw=myyellow}]  table [x=Cell Area, y=Critical Path Slack, col sep=comma] {pgfplot/data/mul/8bit/dw_pareto_output.csv};
    \addplot[blue, style={mark=*, mark size=0.7pt, line width=0.8pt, draw=myorange}]  table [x=Cell Area, y=Critical Path Slack, col sep=comma] {pgfplot/data/mul/8bit/ours_pareto_output.csv};
\coordinate (left) at (rel axis cs:0,1);

\nextgroupplot[minor tick num=0,
xlabel={Area ($\upmu$m$^2$)},
y label style={at={(-0.2,0.5)}},
ylabel near ticks,
legend style={
    draw=none,
	at={(0.45,1.3)},
	nodes={scale=0.75, transform shape},
	anchor=north,
	legend columns=-1,
}
]   
    \addplot[blue, style={mark=*, mark size=0.7pt, line width=0.8pt, draw=mydarkblue}]  table [x=Cell Area, y=Critical Path Slack, col sep=comma]{pgfplot/data/mul/16bit/GOMIL_pareto_output.csv};\addlegendentry{GOMIL\cite{Datapath-2021DATE-Xiao}};
    \addplot[blue, style={mark=*, mark size=0.7pt, line width=0.8pt, draw=myblue}]  table [x=Cell Area, y=Critical Path Slack, col sep=comma] {pgfplot/data/mul/16bit/rl_pareto_output.csv};\addlegendentry{RL-MUL\cite{Datapath-2023DAC-Zuo}};
    \addplot[blue, style={mark=*, mark size=0.7pt, line width=0.8pt, draw=myyellow}]  table [x=Cell Area, y=Critical Path Slack, col sep=comma] {pgfplot/data/mul/16bit/dw_pareto_output.csv};\addlegendentry{Commercial IP};
    \addplot[blue, style={mark=*, mark size=0.7pt, line width=0.8pt, draw=myorange}]  table [x=Cell Area, y=Critical Path Slack, col sep=comma] {pgfplot/data/mul/16bit/ours_pareto_output.csv};\addlegendentry{UFO-MAC};
\coordinate (mid) at (rel axis cs:0.5,1);

\nextgroupplot[minor tick num=0,
xlabel={Area ($\upmu$m$^2$)},
xtick={5000,7500},
y label style={at={(-0.2,0.5)}},
ylabel near ticks,
legend style={
    draw=none,
	at={(0.9,1.2)},
	nodes={scale=0.75, transform shape},
	anchor=north,
	legend columns=-1,
}
]
    \addplot[blue, style={mark=*, mark size=0.7pt, line width=0.8pt, draw=mydarkblue}]  table [x=Cell Area, y=Critical Path Slack, col sep=comma] {pgfplot/data/mul/32bit/gomil_pareto_output.csv};
    \addplot[blue, style={mark=*, mark size=0.7pt, line width=0.8pt, draw=myblue}]  table [x=Cell Area, y=Critical Path Slack, col sep=comma] {pgfplot/data/mul/32bit/rl_new_pareto_output.csv};
    \addplot[blue, style={mark=*, mark size=0.7pt, line width=0.8pt, draw=myyellow}]  table [x=Cell Area, y=Critical Path Slack, col sep=comma] {pgfplot/data/mul/32bit/dw_new_pareto_output.csv};
    \addplot[blue, style={mark=*, mark size=0.7pt, line width=0.8pt, draw=myorange}]  table [x=Cell Area, y=Critical Path Slack, col sep=comma] {pgfplot/data/mul/32bit/ours_new_pareto_output.csv};
\coordinate (right) at (rel axis cs:1,1);

\end{groupplot}
\path (left)--(right) coordinate[midway] (group center);
\path (myplot c2r1.north west|-current bounding box.north)--
coordinate(legendpos)
(myplot c2r1.north west|-current bounding box.north);
\end{tikzpicture}
    \vspace{-0.8cm}
    \caption{Pareto-frontiers of the synthesized results on multipliers. From left to right: 8-bit; 16-bit; 32-bit.}
    \label{fig:results-mul}
\end{figure}
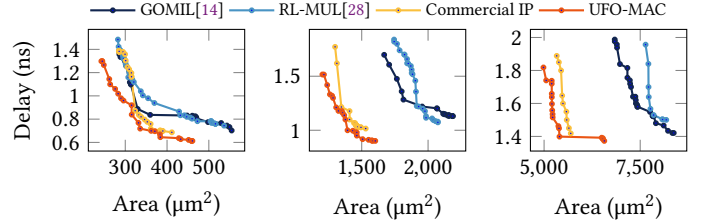

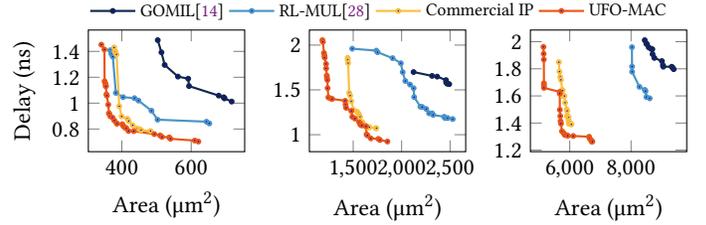
\begin{figure}[!tb]
    \centering

\definecolor{CUHKorange}{RGB}{244,106,18} 
\definecolor{CUHKblue}{RGB}{0,111,190}    
\definecolor{CUHKgreen}{RGB}{0,127,128}   
\definecolor{CUHKmiddle}{RGB}{144,44,144} 
\definecolor{CUHKdark}{RGB}{114,44,114}   
\definecolor{CUHKred}{RGB}{228,46,36}     
\definecolor{CUHKyellow}{RGB}{198,148,34} 

\definecolor{myblue}{RGB}{73,148,196}   
\definecolor{mydarkblue}{RGB}{18,38,79} 
\definecolor{myorange}{RGB}{234,85,20}  
\definecolor{myyellow}{RGB}{250,192,61} 
\definecolor{mypink}{RGB}{252,228,215}  
\definecolor{mygreen}{RGB}{19,138,7}  


\begin{tikzpicture}
\begin{groupplot}[group style={group size= 3 by 1, horizontal sep=0.86cm, group name=myplot}, height=3.2cm, width=3.66cm]
\nextgroupplot[minor tick num=0,
xlabel={Area ($\upmu$m$^2$)},
ylabel={Delay (ns)},
y label style={at={(-0.2,0.5)}},
ylabel near ticks,
legend style={
    draw=none,
	at={(0.5,1.15)},
	nodes={scale=0.75, transform shape},
	anchor=north,
	legend columns=-1,
}
]   
    \addplot[blue, style={mark=*, mark size=0.7pt, line width=0.8pt, draw=mydarkblue}]  table [x=Cell Area, y=Critical Path Slack, col sep=comma] {pgfplot/data/mac/8bit/gomil_pareto_output.csv};
    \addplot[blue, style={mark=*, mark size=0.7pt, line width=0.8pt, draw=myblue}]  table [x=Cell Area, y=Critical Path Slack, col sep=comma] {pgfplot/data/mac/8bit/rl_pareto_output.csv};
    \addplot[blue, style={mark=*, mark size=0.7pt, line width=0.8pt, draw=myyellow}]  table [x=Cell Area, y=Critical Path Slack, col sep=comma] {pgfplot/data/mac/8bit/dw_pareto_output.csv};
    \addplot[blue, style={mark=*, mark size=0.7pt, line width=0.8pt, draw=myorange}]  table [x=Cell Area, y=Critical Path Slack, col sep=comma] {pgfplot/data/mac/8bit/ours_pareto_output.csv};
\coordinate (left) at (rel axis cs:0,1);

\nextgroupplot[minor tick num=0,
xlabel={Area ($\upmu$m$^2$)},
y label style={at={(-0.2,0.5)}},
legend style={
    draw=none,
	at={(0.45,1.3)},
	nodes={scale=0.75, transform shape},
	anchor=north,
	legend columns=-1,
}
]   
    \addplot[blue, style={mark=*, mark size=0.7pt, line width=0.8pt, draw=mydarkblue}]  table [x=Cell Area, y=Critical Path Slack, col sep=comma] {pgfplot/data/mac/16bit/gomil_pareto_output.csv};\addlegendentry{GOMIL\cite{Datapath-2021DATE-Xiao}};
    \addplot[blue, style={mark=*, mark size=0.7pt, line width=0.8pt, draw=myblue}]  table [x=Cell Area, y=Critical Path Slack, col sep=comma] {pgfplot/data/mac/16bit/rl_pareto_output.csv};\addlegendentry{RL-MUL\cite{Datapath-2023DAC-Zuo}};
    \addplot[blue, style={mark=*, mark size=0.7pt, line width=0.8pt, draw=myyellow}]  table [x=Cell Area, y=Critical Path Slack, col sep=comma] {pgfplot/data/mac/16bit/dw_pareto_output.csv};\addlegendentry{Commercial IP};
    \addplot[blue, style={mark=*, mark size=0.7pt, line width=0.8pt, draw=myorange}]  table [x=Cell Area, y=Critical Path Slack, col sep=comma] {pgfplot/data/mac/16bit/ours_pareto_output.csv};\addlegendentry{UFO-MAC};
\coordinate (mid) at (rel axis cs:0.5,1);

\nextgroupplot[minor tick num=0,
xlabel={Area ($\upmu$m$^2$)},
y label style={at={(-0.2,0.5)}},
ylabel near ticks,
legend style={
    draw=none,
	at={(0.9,1.2)},
	nodes={scale=0.75, transform shape},
	anchor=north,
	legend columns=-1,
}
]
    \addplot[blue, style={mark=*, mark size=0.7pt, line width=0.8pt, draw=mydarkblue}]  table [x=Cell Area, y=Critical Path Slack, col sep=comma] {pgfplot/data/mac/32bit/gomil_pareto_output.csv};
    \addplot[blue, style={mark=*, mark size=0.7pt, line width=0.8pt, draw=myblue}]  table [x=Cell Area, y=Critical Path Slack, col sep=comma] {pgfplot/data/mac/32bit/rl_pareto_output.csv};
    \addplot[blue, style={mark=*, mark size=0.7pt, line width=0.8pt, draw=myyellow}]  table [x=Cell Area, y=Critical Path Slack, col sep=comma] {pgfplot/data/mac/32bit/dw_pareto_output.csv};
    \addplot[blue, style={mark=*, mark size=0.7pt, line width=0.8pt, draw=myorange}]  table [x=Cell Area, y=Critical Path Slack, col sep=comma] {pgfplot/data/mac/32bit/ours_pareto_output.csv};
\coordinate (right) at (rel axis cs:1,1);

\end{groupplot}
\path (left)--(right) coordinate[midway] (group center);
\path (myplot c2r1.north west|-current bounding box.north)--
coordinate(legendpos)
(myplot c2r1.north west|-current bounding box.north);
\end{tikzpicture}
    \vspace{-0.8cm}
    \caption{Pareto-frontiers of the synthesized results on MACs. From left to right: 8-bit; 16-bit; 32-bit.}
    \label{fig:results-mac}
\end{figure}

All designs are synthesized by Synopsys Design Compiler (version T-2022.03-SP1)\cite{DesignCompiler} with the NanGate 45nm Open Cell Library\cite{nangate45} and the \texttt{compile\_ultra} command. 
To illustrate the trade-off among the delay, power, and area in various scenarios, we sweep the target delay constraints from $0ns$ to $2ns$ to generate different netlists covering different preferences.

\begin{table*}[!tb]
\caption{FIR filter comparison.}
\vspace{-0.4cm}
\centering
\resizebox{0.98\linewidth}{!}{
\begin{tabular}{|c|c|lccc|lccc|lccc|}
\hline
\multirow{2}{*}{Constraint} & \multirow{2}{*}{Method} & \multicolumn{4}{c|}{8-bit} & \multicolumn{4}{c|}{16-bit} & \multicolumn{4}{c|}{32-bit} \\
\cline{3-14}
& & \multicolumn{1}{c}{Freq (Hz)} & WNS (ns)& Area ($\upmu$m$^2$) & Power (mW) & \multicolumn{1}{c}{Freq (Hz)} & WNS (ns)& Area ($\upmu$m$^2$) & Power (mW) & \multicolumn{1}{c}{Freq (Hz)} & WNS (ns)& Area ($\upmu$m$^2$) & Power (mW) \\
\hline
\multirow{4}{*}{Area-driven} & GOMIL\cite{Datapath-2021DATE-Xiao} & \multicolumn{1}{c|}{\multirow{4}{*}{660M}} & -0.4968 & 2354 & 1.5663 & \multicolumn{1}{c|}{\multirow{4}{*}{500M}} & -0.4990 & 9405 & 8.7474 & \multicolumn{1}{c|}{\multirow{4}{*}{400M}} & -0.4993 & 33804 & 36.584 \\
& RL-MUL\cite{Datapath-2023DAC-Zuo} & \multicolumn{1}{c|}{} & -0.3525 & 2318 & 1.4298 & \multicolumn{1}{c|}{} & \textbf{-0.4989} & 8752 & 8.7020 & \multicolumn{1}{c|}{} & \textbf{-0.5008}& 38022 & 44.264 \\
& Commercial IP & \multicolumn{1}{c|}{} & -0.1805 & 2358 & 1.3137 & \multicolumn{1}{c|}{} & \textbf{-0.4989} & 8397 & 6.9946 & \multicolumn{1}{c|}{} & -0.6533 & 31900 & 35.302 \\
& \textbf{UFO-MAC} & \multicolumn{1}{c|}{} & \textbf{-0.1188} & \textbf{1915} & \textbf{1.0934} & \multicolumn{1}{c|}{} & -0.5707 & \textbf{6429} & \textbf{5.8867} & \multicolumn{1}{c|}{} & -0.5486 & \textbf{29820} & \textbf{32.836} \\
\hline
\multirow{4}{*}{Timing-driven} & GOMIL\cite{Datapath-2021DATE-Xiao} & \multicolumn{1}{c|}{\multirow{4}{*}{2G}} & -0.6287 & 3284 & 2.5342 & \multicolumn{1}{c|}{\multirow{4}{*}{1G}} & -0.6303 & 11112 & 12.004 & \multicolumn{1}{c|}{\multirow{4}{*}{660M}} & -0.5085 & 38167 & 46.405 \\
& RL-MUL\cite{Datapath-2023DAC-Zuo} & \multicolumn{1}{c|}{} & -0.5115 & 3067 & 2.3223 & \multicolumn{1}{c|}{} & -0.4992 & 10572 & 10.872 & \multicolumn{1}{c|}{} & -0.4999 & 38898 & 45.361 \\
& Commercial IP & \multicolumn{1}{c|}{} & -0.5205 & 2919 & 2.0671 & \multicolumn{1}{c|}{} & -0.4477 & 8518 & \textbf{7.3785} & \multicolumn{1}{c|}{} & -0.4994 & 32183 & \textbf{35.715} \\
& \textbf{UFO-MAC} & \multicolumn{1}{c|}{} & \textbf{-0.4893} & \textbf{2733} & \textbf{1.7796} & \multicolumn{1}{c|}{} & \textbf{-0.4277} & \textbf{8394} & 7.4621 & \multicolumn{1}{c|}{} & \textbf{-0.4808} & \textbf{32127} & 35.980 \\
\hline
\multirow{4}{*}{Trade-off} & GOMIL\cite{Datapath-2021DATE-Xiao} & \multicolumn{1}{c|}{\multirow{4}{*}{1G}} & -0.5468 & 2757 & 1.8771 & \multicolumn{1}{c|}{\multirow{4}{*}{660M}} & -0.4662 & 10373 & 10.615 & \multicolumn{1}{c|}{\multirow{4}{*}{500M}} & -0.4266 & 35372 & 40.126 \\
& RL-MUL\cite{Datapath-2023DAC-Zuo} & \multicolumn{1}{c|}{} & -0.2998 & 2718 & 1.9156 & \multicolumn{1}{c|}{} & -0.3976 & 10215 & 10.315 & \multicolumn{1}{c|}{} & -0.5039 & 38245 & 44.211 \\
& Commercial IP & \multicolumn{1}{c|}{} & -0.3486 & 2495 & \textbf{1.4829} & \multicolumn{1}{c|}{} & -0.3493 & 8418 & 7.0109 & \multicolumn{1}{c|}{} & -0.4360 & 31510 & 34.551 \\
& \textbf{UFO-MAC} & \multicolumn{1}{c|}{} & \textbf{-0.2623} & \textbf{2349} & 1.5419 & \multicolumn{1}{c|}{} & \textbf{-0.3137} & \textbf{7658} & \textbf{6.4801} & \multicolumn{1}{c|}{} & \textbf{-0.3883} & \textbf{31366} & \textbf{34.217} \\
\hline
\end{tabular}
}
\label{table:fir-results}
\end{table*}

\begin{table*}[!tb]
\caption{Systolic array comparison.}
\vspace{-0.4cm}
\centering
\resizebox{0.86\linewidth}{!}{
\begin{tabular}{|c|c|lccc|lccc|}
\hline
\multirow{2}{*}{Constraint} & \multirow{2}{*}{Method} & \multicolumn{4}{c|}{8-bit} & \multicolumn{4}{c|}{16-bit} \\
\cline{3-10}
& & \multicolumn{1}{c}{Freq (Hz)} & WNS (ns)& Area ($\upmu$m$^2$) & Power (mW) & \multicolumn{1}{c}{Freq (Hz)} & WNS (ns) & Area ($\upmu$m$^2$) & Power (mW) \\
\hline
\multirow{4}{*}{Area-driven} & GOMIL\cite{Datapath-2021DATE-Xiao} & \multicolumn{1}{c|}{\multirow{4}{*}{660M}} & -0.5102& 168370 & 11.572 & \multicolumn{1}{c|}{\multirow{4}{*}{400M}} & -0.4976& 559985 & 35.918 \\
& RL-MUL\cite{Datapath-2023DAC-Zuo} & \multicolumn{1}{c|}{} & \textbf{-0.4239} & 135659 & 10.207 & \multicolumn{1}{l|}{} & -0.5102& 436095 & 41.480 \\
& Commercial IP & \multicolumn{1}{c|}{} & -0.4684 & 136529 & 10.393 & \multicolumn{1}{l|}{} & -0.4828 & 438526 & 40.506 \\
& \textbf{UFO-MAC} & \multicolumn{1}{c|}{} & -0.4974 & \textbf{125334} & \textbf{9.2475} & \multicolumn{1}{l|}{} & \textbf{-0.4697}& \textbf{401782}& \textbf{35.762}\\
\hline
\multirow{4}{*}{Timing-driven} & GOMIL\cite{Datapath-2021DATE-Xiao} & \multicolumn{1}{c|}{\multirow{4}{*}{2G}} & -0.9827 & 190381 & 12.193 & \multicolumn{1}{c|}{\multirow{4}{*}{1G}} & -0.9854 & 662801 & 44.912 \\
& RL-MUL\cite{Datapath-2023DAC-Zuo} & \multicolumn{1}{l|}{} & -0.7077 & 172810 & 11.873 & \multicolumn{1}{l|}{} & -0.5856 & 609563 & 44.275 \\
& Commercial IP & \multicolumn{1}{l|}{} & -0.6053 & 144137 & 11.357 & \multicolumn{1}{l|}{} & -0.3375 & \textbf{467621} & 45.221 \\
& \textbf{UFO-MAC} & \multicolumn{1}{l|}{} & \textbf{-0.5946} & \textbf{138316} & \textbf{10.787} & \multicolumn{1}{l|}{} & \textbf{-0.1994} & 533072 & \textbf{40.164} \\
\hline
\multirow{4}{*}{Trade-off} & GOMIL\cite{Datapath-2021DATE-Xiao} & \multicolumn{1}{c|}{\multirow{4}{*}{1G}} & -0.6842 & 178874 & 11.175 & \multicolumn{1}{c|}{\multirow{4}{*}{660M}} & -0.6611 & 611143 & 41.651 \\
& RL-MUL\cite{Datapath-2023DAC-Zuo} & \multicolumn{1}{l|}{} & -0.6955 & 141754 & 10.892 & \multicolumn{1}{l|}{} & -0.0981 & 564192 & 43.515 \\
& Commercial IP & \multicolumn{1}{l|}{} & -0.6941 & 141905 & 10.831 & \multicolumn{1}{l|}{} & -0.0999 & 458647 & 45.077 \\
& \textbf{UFO-MAC} & \multicolumn{1}{l|}{} & \textbf{-0.6785} & \textbf{131083} & \textbf{9.5777} & \multicolumn{1}{l|}{} & \textbf{-0.0182} & \textbf{449184} & \textbf{36.205} \\
\hline
\end{tabular}
}
\label{table:pe-results}
\end{table*}
 
\subsection{Multiplier and MAC Comparison}
Comparisons of compressor trees in \Cref{fig:results-ct} include only RL-MUL and commercial IP, as GOMIL's compressor tree is merged into its RTL and cannot be exactly decoupled. 
The results demonstrate that UFO-MAC outperforms all baselines. 
Multiplier results in~\Cref{fig:results-mul} reveal that UFO-MAC provides Pareto-optimal performance, with improvements up to 14.9\% in area and 11.3\% in delay compared to commercial multipliers. 
The comprehensive design space exploration including compressor assignment, interconnection order and non-uniform CPA optimization contribute significantly to these improvements over GOMIL and RL-MUL.
While GOMIL focuses only on optimizing the area of the compressor tree, resulting in sub-optimal delays due to neglect of stage and interconnect considerations, it also lacks area efficiency due to its CPA optimization objectives centered solely on the logic level. 
RL-MUL may suffer from scalability issues, especially in larger bit-width scenarios.
MAC results in \Cref{fig:results-mac} confirm that UFO-MAC achieves up to 18.1\% reduction in area and 13.9\% in delay compared to commercial MACs. 
The fused MAC architecture, which merges the accumulator into the partial product generation, offers substantial area and delay savings by eliminating an extra adder stage.

\subsection{Implementation in Functional Modules}
To further validate the performance advantages of our framework in larger-scale designs, we integrated the multipliers and MACs from all approaches into more complex functional modules. 
Specifically, multipliers are incorporated into 5-stage finite impulse response (FIR) filters, commonly utilized in signal processing applications. 
MACs are applied to the implementation of two systolic array designs that are commonly used in AI chips. 
Both designs have $16\times 16$ processing elements and the bit width is 8-bit and 16-bit, respectively\footnote{An optimized 32-bit systolic array implementation is not available. Hence no experiments were conducted on it.}. 
These designs are synthesized under various clock frequency constraints to assess area, timing, and trade-off scenarios.
Results for the FIR filters are detailed in \Cref{table:fir-results}, and those for systolic arrays in \Cref{table:pe-results}.
It can be seen that when applying the obtained multipliers and MACs to larger functional modules implementation, the improvement on delay, power, and area still persists. 

\section{Conclusion}
This work has introduced UFO-MAC, a unified framework aimed at enhancing the optimization of high-performance multipliers and multiply-accumulators. Through the implementation of an optimal compressor tree and then the refinement of stage assignment along with interconnection orders using ILP, coupled with the strategic utilization of the non-uniform arrival profile of carry propagation adders (CPA), UFO-MAC demonstrably surpasses both contemporary benchmarks and commercial tools in performance. 
Experimental validation within FIR filter and systolic array configurations underscores the framework's capability to significantly reduce area and delay, thereby achieving substantial performance improvements. 
Future efforts may explore extending UFO-MAC's methodologies to floating-point multipliers and broader applications, such as datapath designs within Processing Element (PE) arrays, enhancing its utility in increasingly complex computing environments.
\section*{Acknowledgments}
This work is supported in part by the Guangzhou-HKUST(GZ) Joint Funding Program (No. 2023A03J0155) and Guangzhou Municipal Science and Technology Project (Municipal Key Laboratory Construction Project, Grant No.2023A03J0013).
\clearpage
{
\bibliographystyle{IEEEtran}
\bibliography{ref/Top-sim, ref/Datapath}

\begin{thebibliography}{10}
\providecommand{\url}[1]{#1}
\csname url@samestyle\endcsname
\providecommand{\newblock}{\relax}
\providecommand{\bibinfo}[2]{#2}
\providecommand{\BIBentrySTDinterwordspacing}{\spaceskip=0pt\relax}
\providecommand{\BIBentryALTinterwordstretchfactor}{4}
\providecommand{\BIBentryALTinterwordspacing}{\spaceskip=\fontdimen2\font plus
\BIBentryALTinterwordstretchfactor\fontdimen3\font minus
  \fontdimen4\font\relax}
\providecommand{\BIBforeignlanguage}[2]{{%
\expandafter\ifx\csname l@#1\endcsname\relax
\typeout{** WARNING: IEEEtran.bst: No hyphenation pattern has been}%
\typeout{** loaded for the language `#1'. Using the pattern for}%
\typeout{** the default language instead.}%
\else
\language=\csname l@#1\endcsname
\fi
#2}}
\providecommand{\BIBdecl}{\relax}
\BIBdecl

\bibitem{Datapath-1964TC-Wallace}
C.~S. Wallace, ``A suggestion for a fast multiplier,'' \emph{IEEE Transactions
  on Electronic Computers}, vol. EC-13, no.~1, pp. 14--17, 1964.

\bibitem{Datapath-1983ARITH-Dadda}
L.~Dadda, ``Some schemes for fast serial input multipliers,'' in \emph{1983
  IEEE 6th Symposium on Computer Arithmetic (ARITH)}, 1983, pp. 52--59.

\bibitem{Datapath-1993ASAP-Bickerstaff}
K.~Bickerstaff, M.~Schulte, and E.~Swartzlander, ``Reduced area multipliers,''
  in \emph{Proceedings of International Conference on Application Specific
  Array Processors (ASAP '93)}, 1993, pp. 478--489.

\bibitem{Datapath-1993TVLSI-Fadavi-Ardekani}
J.~Fadavi-Ardekani, ``M*n booth encoded multiplier generator using optimized
  wallace trees,'' \emph{IEEE TVLSI}, vol.~1, no.~2, pp. 120--125, June 1993.

\bibitem{Datapath-2005ISCAS-Itoh}
N.~Itoh, Y.~Tsukamoto, T.~Shibagaki, K.~Nii, H.~Takata, and H.~Makino, ``A
  32/spl times/24-bit multiplier-accumulator with advanced rectangular styled
  wallace-tree structure,'' in \emph{Proc.~ISCAS}, 2005, pp. 73--76 Vol. 1.

\bibitem{Datapath-2014ATC-Luu}
X.-V. Luu, T.-T. Hoang, T.-T. Bui, and A.-V. Dinh-Duc, ``A high-speed unsigned
  32-bit multiplier based on booth-encoder and wallace-tree modifications,'' in
  \emph{2014 International Conference on Advanced Technologies for
  Communications (ATC 2014)}, 2014, pp. 739--744.

\bibitem{Datapath-1996TC-Oklobdzija}
V.~Oklobdzija, D.~Villeger, and S.~Liu, ``A method for speed optimized partial
  product reduction and generation of fast parallel multipliers using an
  algorithmic approach,'' \emph{IEEE Transactions on Computers}, vol.~45,
  no.~3, pp. 294--306, 1996.

\bibitem{Datapath-1995ARITH-Martel}
C.~Martel, V.~Oklobdzija, R.~Ravi, and P.~Stelling, ``Design strategies for
  optimal multiplier circuits,'' in \emph{Proceedings of the 12th Symposium on
  Computer Arithmetic}, 1995, pp. 42--49.

\bibitem{Datapath-1998TC-Stelling}
P.~Stelling, C.~Martel, V.~Oklobdzija, and R.~Ravi, ``Optimal circuits for
  parallel multipliers,'' \emph{IEEE Transactions on Computers}, vol.~47,
  no.~3, pp. 273--285, 1998.

\bibitem{Datapath-2008DATE-Parandeh-Afshar}
H.~Parandeh-Afshar, P.~Brisk, and P.~Ienne, ``Improving synthesis of compressor
  trees on fpgas via integer linear programming,'' in \emph{2008 Design,
  Automation and Test in Europe}, 2008, pp. 1256--1261.

\bibitem{Datapath-2014FPL-Kumm}
M.~Kumm and P.~Zipf, ``Pipelined compressor tree optimization using integer
  linear programming,'' in \emph{Proc.~FPL}, 2014, pp. 1--8.

\bibitem{Datapath-2017ARITH-Kumm}
M.~Kumm, J.~Kappauf, M.~Istoan, and P.~Zipf, ``Resource optimal design of large
  multipliers for fpgas,'' in \emph{2017 IEEE 24th Symposium on Computer
  Arithmetic (ARITH)}, 2017, pp. 131--138.

\bibitem{Datapath-2018TC-Kumm}
M.~Kumm and J.~Kappauf, ``Advanced compressor tree synthesis for fpgas,''
  \emph{IEEE Transactions on Computers}, vol.~67, no.~8, pp. 1078--1091, 2018.

\bibitem{Datapath-2021DATE-Xiao}
W.~Xiao, W.~Qian, and W.~Liu, ``Gomil: Global optimization of multiplier by
  integer linear programming,'' in \emph{2021 Design, Automation and Test in
  Europe Conference and Exhibition (DATE)}, 2021, pp. 374--379.

\bibitem{Datapath-1960TC-Sklansky}
J.~Sklansky, ``Conditional-sum addition logic,'' \emph{IRE Transactions on
  Electronic Computers}, vol. EC-9, no.~2, pp. 226--231, 1960.

\bibitem{Datapath-1973TC-Kogge}
P.~M. Kogge and H.~S. Stone, ``A parallel algorithm for the efficient solution
  of a general class of recurrence equations,'' \emph{IEEE Transactions on
  Computers}, vol. C-22, no.~8, pp. 786--793, 1973.

\bibitem{Datapath-1982TC-Brent}
Brent and Kung, ``A regular layout for parallel adders,'' \emph{IEEE
  Transactions on Computers}, vol. C-31, no.~3, pp. 260--264, 1982.

\bibitem{Datapath-1990DAC-Fishburn}
J.~Fishburn, ``A depth-decreasing heuristic for combinational logic; or how to
  convert a ripple-carry adder into a carry-lookahead adder or anything
  in-between,'' in \emph{Proc.~DAC}, 1990, pp. 361--364.

\bibitem{1996-IWLAS-Zimmermann}
R.~Zimmermann, ``Non-heuristic optimization and synthesis of parallel-prefix
  adders,'' in \emph{International Workshop on Logic and Architecture
  Synthesis}, 1996.

\bibitem{Datapath-2002ISPAN-Fishburn}
Y.-C. Lin and J.-W. Hsiao, ``A new approach to constructing optimal prefix
  circuits with small depth,'' in \emph{Proceedings International Symposium on
  Parallel Architectures, Algorithms and Networks. I-SPAN'02}, 2002, pp.
  99--104.

\bibitem{Datapath-2007ASPDAC-Liu}
J.~Liu, Y.~Zhu, H.~Zhu, C.-K. Cheng, and J.~Lillis, ``Optimum prefix adders in
  a comprehensive area, timing and power design space,'' in
  \emph{Proc.~ASPDAC}, 2007, pp. 609--615.

\bibitem{Datapath-2013DAC-Roy}
S.~Roy, M.~Choudhury, R.~Puri, and D.~Z. Pan, ``Towards optimal
  performance-area trade-off in adders by synthesis of parallel prefix
  structures,'' in \emph{Proc.~DAC}, 2013, pp. 1--8.

\bibitem{Datapath-2014TCAD-Roy}
------, ``Towards optimal performance-area trade-off in adders by synthesis of
  parallel prefix structures,'' \emph{IEEE Transactions on Computer-Aided
  Design of Integrated Circuits and Systems}, vol.~33, no.~10, pp. 1517--1530,
  2014.

\bibitem{Datapath-2016TACD-Roy}
------, ``Polynomial time algorithm for area and power efficient adder
  synthesis in high-performance designs,'' \emph{IEEE Transactions on
  Computer-Aided Design of Integrated Circuits and Systems}, vol.~35, no.~5,
  pp. 820--831, 2016.

\bibitem{DSE-2022TCAD-Geng}
H.~Geng, Y.~Ma, Q.~Xu, J.~Miao, S.~Roy, and B.~Yu, ``High-speed adder design
  space exploration via graph neural processes,'' \emph{IEEE TCAD}, vol.~41,
  no.~8, pp. 2657--2670, 2022.

\bibitem{DSE-2019TCAD-MA}
Y.~Ma, S.~Roy, J.~Miao, J.~Chen, and B.~Yu, ``Cross-layer optimization for high
  speed adders: A pareto driven machine learning approach,'' \emph{IEEE TCAD},
  vol.~38, no.~12, pp. 2298--2311, 2019.

\bibitem{RL-2021DAC-Roy}
R.~Roy, J.~Raiman, N.~Kant, I.~Elkin, R.~Kirby, M.~Siu, S.~Oberman, S.~Godil,
  and B.~Catanzaro, ``Prefixrl: Optimization of parallel prefix circuits using
  deep reinforcement learning,'' in \emph{Proc.~DAC}, 2021, pp. 853--858.

\bibitem{Datapath-2023DAC-Zuo}
D.~Zuo, Y.~Ouyang, and Y.~Ma, ``Rl-mul: Multiplier design optimization with
  deep reinforcement learning,'' in \emph{2023 60th ACM/IEEE Design Automation
  Conference (DAC)}, 2023, pp. 1--6.

\bibitem{Datapath-1995Asilomar-Stelling}
P.~F. Stelling and V.~Oklobdzija, ``Design strategies for the final adder in a
  parallel multiplier,'' in \emph{Proceedings of the 29th Asilomar Conference
  on Signals, Systems and Computers (2-Volume Set)}, ser. ASILOMAR '95.\hskip
  1em plus 0.5em minus 0.4em\relax IEEE Computer Society, 1995, p. 591.

\bibitem{Datapath-1995TVLSI-Oklobdzija}
V.~Oklobdzija and D.~Villeger, ``Improving multiplier design by using improved
  column compression tree and optimized final adder in cmos technology,''
  \emph{IEEE Transactions on Very Large Scale Integration (VLSI) Systems},
  vol.~3, no.~2, pp. 292--301, 1995.

\bibitem{Datapath-2012TODAES-Kim}
\BIBentryALTinterwordspacing
Y.~Kim, S.~Kwak, and T.~Kim, ``Synthesis of adaptable hybrid adders for area
  optimization under timing constraint,'' \emph{ACM Trans. Des. Autom.
  Electron. Syst.}, vol.~17, no.~4, oct 2012. [Online]. Available:
  \url{https://doi.org/10.1145/2348839.2348847}
\BIBentrySTDinterwordspacing

\bibitem{Datapath-2007GLSVLSI-Matsunaga}
\BIBentryALTinterwordspacing
T.~Matsunaga and Y.~Matsunaga, ``Area minimization algorithm for parallel
  prefix adders under bitwise delay constraints,'' in \emph{Proceedings of the
  17th ACM Great Lakes Symposium on VLSI}, ser. GLSVLSI '07.\hskip 1em plus
  0.5em minus 0.4em\relax New York, NY, USA: Association for Computing
  Machinery, 2007, p. 435–440. [Online]. Available:
  \url{https://doi.org/10.1145/1228784.1228886}
\BIBentrySTDinterwordspacing

\bibitem{Datapath-2003Asilomar-Harris}
D.~Harris and I.~Sutherland, ``Logical effort of carry propagate adders,'' in
  \emph{The Thrity-Seventh Asilomar Conference on Signals, Systems and
  Computers, 2003}, vol.~1, 2003, pp. 873--878 Vol.1.

\bibitem{Datapath-1986JA-Snir}
\BIBentryALTinterwordspacing
M.~Snir, ``Depth-size trade-offs for parallel prefix computation,'' \emph{J.
  Algorithms}, vol.~7, no.~2, p. 185–201, jun 1986. [Online]. Available:
  \url{https://doi.org/10.1016/0196-6774(86)90003-9}
\BIBentrySTDinterwordspacing

\bibitem{Berkerly-ABC}
{Berkeley Logic Synthesis and Verification Group}, ``{ABC: A System for
  Sequential Synthesis and Verification},''
  \url{https://people.eecs.berkeley.edu/~alanmi/abc/}.

\bibitem{Gurobi}
{Gurobi Optimization, LLC}, ``{GUROBI OPTIMIZER},''
  \url{https://www.gurobi.com}.

\bibitem{DesignCompiler}
{Synopsys, Inc.}, ``{Design Compiler},''
  \url{https://www.synopsys.com/implementation-and-signoff/rtl-synthesis-test/dc-ultra.html}.

\bibitem{nangate45}
\BIBentryALTinterwordspacing
{Nangate Inc.}, ``{Open Cell Library v2008\_10 SP1},'' 2008. [Online].
  Available: \url{http://www.nangate.com/openlibrary/}
\BIBentrySTDinterwordspacing

\end{thebibliography}
}
\end{document}